\documentclass[final,11p,times,twocolumn,floatfix,prd,authoryear,
nofootinbib,aps,eqsecnum,superscriptaddress,preprintnumbers]{revtex4-2}

\usepackage{bm}
\usepackage{newtxmath,newtxtext}

\usepackage{graphicx}

\usepackage{textcomp}
\usepackage{hyperref}
\hypersetup{colorlinks=true,linkcolor=red,urlcolor=blue,	citecolor=red}

\usepackage{xcolor}
\usepackage{appendix}
\newcommand{\bq}{\begin{equation}}
	\newcommand{\eq}{\end{equation}}
\newcommand{\bqn}{\begin{eqnarray}}
	\newcommand{\eqn}{\end{eqnarray}}
\newcommand{\nb}{\nonumber}
\newcommand{\lb}{\label}

\newcommand{\ct}{\ensuremath{c_{\rm T}^{2}}}

\begin{document}

% \preprint{YITP-21-11, IPMU21-0011}

\title{Interacting bosonic dark energy and fermionic dark matter in Einstein scalar Gauss-Bonnet gravity}

% \title{Interacting Models of Dark Energy and Fermionic Dark Matter of particle physics in Einstein Scalar Gauss-Bonnet Gravity}

% \title{Interacting Models of Dark Energy and Dark Matter in Einstein scalar Gauss-Bonnet Gravity with fermionic dark matter field}

\author{Simran Arora}
\email{arora.simran@yukawa.kyoto-u.ac.jp}
\affiliation{Center for Gravitational Physics and Quantum Information, Yukawa Institute for Theoretical Physics, Kyoto University, 606-8502, Kyoto, Japan}

\author{Saddam Hussain}
\email{saddamh@zjut.edu.cn}
\affiliation{Institute for Theoretical Physics and Cosmology, Zhejiang University of Technology, Hangzhou 310023, China}

\author{Benjamin Rose}
\email{Ben$\_$Rose@baylor.edu}
\affiliation{Department of Physics and Astronomy, Baylor University, Waco, TX 76798-7316, USA}

\author{Anzhong Wang}
\email{anzhong$\_$wang@baylor.edu}
\affiliation{GCAP-CASPER, Department of Physics and Astronomy, Baylor University, Waco, TX 76798-7316, USA}

\begin{abstract}

We explore a cosmological framework in which a Gauss-Bonnet (GB) coupled scalar field, acting as dark energy, interacts with a fermionic dark matter field through a coupling obtained from the point of view of particle physics. This setup is inspired by string/M-theory, and two representative scalar field potentials are investigated: exponential and power-law. A distinctive feature of the GB-coupled models is their potential to alter the propagation speed of gravitational waves (GWs), a property with significant implications in light of recent multi-messenger astrophysical observations. To account for this, we analyze models under two scenarios: one where the GW speed differs from that of light and the other where they are equal, but all consistent with current observational constraints. The dynamical evolution of the system is investigated by reformulating the field equations into an autonomous dynamical system, enabling a detailed analysis of the Universe's long-term behavior, including the radiation-, matter- and dark energy-dominated epochs. We constrain the model parameters using a broad set of recent observational data, including mock high-redshift measurements from the Roman Space Telescope. Our findings indicate that both potentials yield cosmologies that are in excellent agreement with current data, closely tracking the expansion history predicted by the standard \(\Lambda\)CDM model, while still allowing room for subtle deviations that could be tested by future observations.\\

\end{abstract}

\maketitle

\section{Introduction}

The discovery of the accelerated expansion of the Universe at late times~\cite{SupernovaSearchTeam:1998fmf,SupernovaCosmologyProject:1998vns} has significantly challenged our understanding of the fundamental physical laws and fields that govern cosmic evolution. Within the framework of a statistically homogeneous and isotropic universe, this acceleration is most commonly attributed to a hypothetical fluid with negative pressure, referred to as dark energy, % (DE), 
typically modeled by a positive cosmological constant \(\Lambda\), characterized by an equation of state \(w_\Lambda = -1\). However, theoretical challenges associated with the cosmological constant, including its striking discrepancy with predictions from quantum field theory~\cite{Weinberg:1988cp} and the requirement of finely tuned initial conditions~\cite{Zlatev:1998tr}, have motivated the search for more fundamental mechanisms underlying the observed late-time acceleration~\cite{Joyce:2014kja}. Despite these efforts, no alternative model has gained widespread acceptance than the \(\Lambda\)CDM framework, which still remains the prevailing cosmological paradigm, assuming the validity of the cosmological constant hypothesis.

Over the past decade, increasingly precise cosmological observations have revealed significant tensions within the \(\Lambda\)CDM framework. Most notably, there exists around \(5\sigma\) discrepancy between the Hubble constant inferred from cosmic microwave background (CMB) measurements by the \textit{Planck} satellite~\cite{Planck:2018vyg} and the local distance-ladder estimations reported by the SH0ES collaboration~\cite{riess2022comprehensive}. There also exist tensions between the galactic scale matter clustering inferred from Planck and the one derived from galaxy surveys and weak gravitational lensing observations \cite{DiValentino:2020vvd,DES:2021bvc,DES:2021wwk}. These persistent discrepancies strongly suggest that the foundational assumptions of the standard cosmological model may require revision or replacement. Recently, the Dark Energy Spectroscopic Instrument (DESI) collaboration reported a \(3.9\sigma\) indication of dynamical dark energy in its first data release (DR1), which was further strengthened to \(4.2\sigma\) in the second data release (DR2)~\cite{DESI:2024mwx,DESI:2025zgx}. This result has generated significant interest and sparked active discussions regarding potential signatures of new physics beyond the standard \(\Lambda\)CDM paradigm.

A wide variety of theoretical models have been proposed to explain the late-time acceleration of the Universe. Among the most widely studied are scalar field models, such as  quintessence~\cite{Peebles:2002gy,Berghaus:2024kra,Kritpetch:2024rgi}, where the acceleration is driven by the potential energy of a slowly evolving scalar field. Another class involves kinetically driven scenarios, known as k-essence~\cite{Hussain:2024qrd,Armendariz-Picon:2000ulo,Scherrer:2004au,Hussain:2022osn,Bhattacharya:2022wzu}, in which the dynamics of the scalar field are governed primarily by non-standard kinetic terms. In addition, several non-canonical scalar field models have been explored, including tachyonic and ghost fields ~\cite{Bagla:2002yn,Khoeini-Moghaddam:2018znw,Cai:2010uf,Hussain:2022dhp}. 

Another promising direction involves modifying the gravitational sector itself. In this context, modified gravity theories~\cite{Sotiriou:2008rp,Nojiri:2006ri,Sokoliuk:2023ccw,Gorji:2020bfl,DeFelice:2020eju,Arora:2022dti,Gadbail:2024syv} and scalar–tensor formulations~\cite{Kobayashi:2019hrl,Elizalde:2023rds,Odintsov:2025kyw,Hussain:2023kwk} extend the geometric structure of General Relativity (GR) and offer alternative explanations for the observed cosmic acceleration without introducing additional exotic matter components. 
Within this broad class of theories, Einstein–scalar–Gauss–Bonnet (EsGB) gravity has attracted considerable attention. The Gauss–Bonnet term was originally introduced by Lanczos~\cite{Lanczos:1938sf} and later generalized within the Lovelock framework~\cite{Lovelock:1971yv}. A key feature of this construction is that it allows modifications of gravity while preserving second-order field equations. This property is particularly important because higher-derivative theories often suffer from the Ostrogradsky instability~\cite{Ostrogradsky:1850fid}. Interestingly, EsGB gravity also emerges naturally as an effective low-energy limit of string and M-theory~\cite{Gross:1986mw,Bento:1995qc,Ferrara:1996hh,Antoniadis:1997eg}. As a result, it provides a theoretically well-motivated framework in which higher-curvature corrections to GR can be studied without introducing pathological degrees of freedom. For a more detailed discussion of these aspects, see Ref.~\cite{Wang:2017brl} and the references therein.

The EsGB theory extends the Einstein–Hilbert action by introducing a scalar field coupled to the Gauss–Bonnet term, a specific combination of quadratic curvature invariants. This term emerges naturally in the low-energy effective action of string theory and captures higher-curvature corrections without introducing ghost instabilities \cite{Nojiri:2005vv,Tsujikawa:2006ph,Cognola:2006sp}. A particularly appealing feature of the Gauss–Bonnet combination is that it is the only curvature-squared correction that preserves the second-order nature of the gravitational field equations. This property ensures that the theory avoids the pathological instabilities that often appear in higher-derivative gravity models, while maintaining the consistency and predictive power of the underlying framework~\cite{Odintsov:2019clh,Antoniou:2017hxj,Julie:2019sab,East:2022rqi,Millano:2023gkt,Granda:2014zea,Zhang:2020pko}. From both theoretical and phenomenological perspectives, EsGB gravity offers a compelling framework to explain cosmic acceleration through dynamical scalar-curvature couplings, rather than a static cosmological constant. Henceforth, these models remain under active investigations, with ongoing efforts aimed at constraining them via astrophysical observations and assessing their consistency with cosmological data.

In this work, we explore a framework in which both dark matter and dark energy are described as fields from the point of view of particle physics rather than as fluids. This framework allows us to write down naturally the interactions between the two dark components. 
Interacting models between the two dark components have garnered great interest for their potential to address the Hubble tension and the coincidence problem~\cite{Wang:2016lxa,Wang:2024vmw,Benisty:2024lmj,DiValentino:2019jae,DiValentino:2021izs,Yang:2018euj,Wang:2014iua}. In particular, we consider a framework in which dark energy is modeled by a scalar field, while dark matter is described by a fermionic field in Gauss-Bonnet theory. The latter provides a particle-based description of dark matter whose energy density and pressure enter the Friedmann equations and therefore influence the background cosmological dynamics. Fermionic fields have been extensively studied in this context and have been shown to admit consistent cosmological solutions capable of describing inflationary phases, dark energy–like behavior, or effective dark matter components, depending on their self-interactions and couplings~\cite{Chakraborty:2025syu,Samojeden:2010rs,Ribas:2016ulz,Benisty:2019pxb,Grams:2014woa,Farrar:2003uw,Bean:2008ac,Feng:2010gw}.

In our setup, a non-gravitational interaction between the scalar and fermionic fields is introduced at the action level via the term \(\mathcal{L}_{\text{int}}(\phi, \psi)\), where $\phi$ and $\psi$ denote respectively the GB couple scalar field and the fermionic scalar field. This term facilitates a direct exchange of energy and momentum between the dark components, leading to modified equations of motion that depart from the standard non-interacting scenario significantly. 
%Although the interaction is introduced phenomenologically at the action level, it remains fully consistent with the variational principle and leads to well-defined field equations. 
Crucially, this formulation naturally constrains the form of the interaction, avoiding the arbitrariness often found in purely phenomenological models. In particular, this stands in contrast to fluid-based interacting dark sector approaches, where interactions are typically imposed through an ad hoc source term in the continuity equations~\cite{Wang:2014iua,Benisty:2024lmj,Bolotin:2013jpa,Chatterjee:2021ijw,Hussain:2024jdt}.

Furthermore, due to the coupling between the scalar field and the Gauss-Bonnet term, the theory introduces higher-curvature corrections that modify the gravitational dynamics. A key prediction of such models is a deviation in the propagation speed of gravitational waves (GWs), which has been tightly constrained by recent multi-messenger observations~\cite{LIGOScientific:2017zic,TerenteDiaz:2023iqk,Ezquiaga:2017ekz}. In particular, the joint detection of GWs and gamma-ray bursts from the binary neutron star merger GW170817 imposes a stringent bound on the GW speed
$$
\left|c_T^2 - 1\right| < 5 \times 10^{-16}, 
$$
where \( c_T \) denotes the GW speed in natural units. In light of this, we consider two scenarios in our analysis: one with \( c_T^2 \neq 1 \) and the other with  \( c_T^2 = 1 \), but in both scenarios we assume that the above observational constraint is satisfied.

% In this work, our objective is not to perform a full analytical dynamical-systems classification of all fixed points through Jacobian eigenvalue analysis. Instead, we construct the autonomous system using phase-space variables and investigate its behavior numerically. Due to the presence of additional degrees of freedom in the model, the dimensionality of the phase space exceeds three, which renders a complete analytical eigenvalue analysis technically cumbersome and not particularly illuminating for observational purposes. To assess the stability of the system, we numerically evolve the autonomous equations from the deep radiation epoch ($N \simeq -20$) to the future era ($N > 0$). A viable region of the parameter space is identified by imposing the following conditions: the dynamical variables remain bounded and physically meaningful throughout the evolution. The system approaches a late-time accelerated attractor. At the attractor, the scalar-field density parameter satisfies $\Omega_{\phi} > 0.5$ and the effective equation of state obeys $w_{\mathrm{eff}} < -0.3$. Further in the present analysis, the dynamics of the coupled system are investigated using focusing on models with exponential and power-law scalar potentials. 
Within the spatially flat FLRW background, we derive the equations of motion and relevant cosmological quantities. To constrain the model parameters, we employ a comprehensive set of observational data, including CMB, Hubble parameter measurements, Type Ia Supernovae (Pantheon+ and DESY5), DESI Baryon Acoustic Oscillation data (DR2), and mock data from the Roman Space Telescope.

The structure of the paper is as follows:  Sec.~\ref{sec:theoretical_framework} presents the theoretical framework, including the background equations and cosmological quantities. Sec.~\ref{dynamics} outlines the dynamics of the interacting models. Sec.~\ref{sec:data} describes the observational data sets used for parameter constraints. Sec.~\ref {results} discusses the results, and finally Sec.~\ref {conc} concludes with a summary of our main findings. We also include an appendix, Appendix A, in which we investigate the autonomous systems numerically and show explicitly the robustness of the initial conditions, adopted in the models considered in this paper. 
 
 It should be noted that in this work our objective is not to perform a full analytical dynamical-systems classification of all fixed points through Jacobian eigenvalue analysis. Instead, we construct the autonomous system using phase-space variables and investigate its behavior numerically. Due to the presence of additional degrees of freedom in the model, the dimensionality of the phase space exceeds three, which renders a complete analytical eigenvalue analysis technically cumbersome and not particularly illuminating for observational purposes. To assess the stability of the system, we numerically evolve the autonomous equations from the deep radiation epoch ($N \simeq -20$) to the future era ($N > 0$). A viable region of the parameter space is identified by imposing the following conditions: the dynamical variables remain bounded and physically meaningful throughout the evolution. The system approaches a late-time accelerated attractor. At the attractor, the scalar-field density parameter satisfies $\Omega_{\phi} > 0.5$ and the effective equation of state obeys $w_{\mathrm{eff}} < -0.3$, as shown explicitly in Appendix A.
% Further in the present analysis, the dynamics of the coupled system are investigated using focusing on models with exponential and power-law scalar potentials. Within the spatially flat FLRW background, we derive the equations of motion and relevant cosmological quantities. To constrain the model parameters, we employ a comprehensive set of observational data, including CMB, Hubble parameter measurements, Type Ia Supernovae (Pantheon+ and DESY5), DESI Baryon Acoustic Oscillation data (DR2), and mock data from the Roman Space Telescope.}

\section{Coupled Einstein-scalar-Gauss-Bonnet gravity}
\label{sec:theoretical_framework}

In this section, let us consider a framework in which dark energy and dark matter are described respectively by a bosonic and a fermionic scalar field, interacting through an effective Lagrangian. 

\subsection{General Framework}

We formulate this scenario within the EsGB gravity framework \cite{Zhang:2020pko,Cicoli:2023opf,Fernandes:2022zrq}, described by the action  
\begin{equation}
    S_{\text{EsGB}} = \int d^4x \sqrt{-g} \left[ \frac{R}{2\kappa^2} + f(\phi) \mathcal{G} + \mathcal{L}_{\phi} + \mathcal{L}_{\psi} + \mathcal{L}_{\text{int}} + \mathcal{L}_{m} \right],
    \label{eq:EsGB_action}
\end{equation}
where \( \kappa \equiv 8\pi G/c^4 \), \( c \) is the speed of light in vacuum, and \( g \equiv \det(g_{\mu \nu}) \). The Ricci scalar is denoted by \( R \), and the Gauss-Bonnet invariant is defined as:
\begin{equation}
    \mathcal{G} = R^2 - 4R_{\mu\nu}R^{\mu\nu} + R_{\mu\nu\rho\sigma}R^{\mu\nu\rho\sigma}.
\end{equation}
Throughout this work, we adopt natural units by setting \( \kappa = 1 \). The scalar field \( \phi \), associated with dark energy, couples non-minimally to the GB term via a general coupling function \( f(\phi) \). All other components of the Universe are assumed to interact minimally with the metric.
%and solely through gravitational interactions. 

Building on this framework, we incorporate the dark matter sector as a non-relativistic Dirac fermion, represented by the field \( \psi \). In addition to its gravitational interaction with other components of matter fields in the Universe, the fermion couples directly to the scalar field $\phi$ through a Yukawa-type interaction, introduced via the interaction Lagrangian \( \mathcal{L}_{\text{int}}(\phi, \psi)\) \cite{Micheletti:2010cm,Micheletti:2010cm,Ribas:2016ulz,McDonough:2021pdg}. The complete dynamics are governed by the following individual Lagrangian components:
\begin{align}
    \mathcal{L}_{\phi} &= -\frac{1}{2} \nabla_\mu \phi \nabla^\mu \phi - V(\phi), \nonumber \\
    \mathcal{L}_{\psi} &= \frac{i}{2} \left[ \overline{\psi} \Gamma^\mu \nabla_\mu \psi - (\nabla_\mu \overline{\psi}) \Gamma^\mu \psi \right] - W(\overline{\psi}, \psi), \nonumber \\
    \mathcal{L}_{\text{int}} &= - M(\phi) \overline{\psi} \psi, \nonumber
    \\
    \mathcal{L}_{\text{m}} &= \mathcal{L}_{\text{m}}(g_{\mu\nu}, \psi_m), 
    \label{eq:Lint}
\end{align}
where \( V(\phi) \) is the scalar potential, \( M(\phi) \) is a field-dependent mass function mediating the coupling between \( \phi \) and \( \psi \), and \( W(\overline{\psi}, \psi) \) accounts for possible fermionic self-interactions. Here, \( \Gamma^\mu = e^\mu{}_a \gamma^a \) are the curved-space gamma matrices, constructed from the vierbeins \( e^\mu{}_a \) and the flat-space Dirac matrices \( \gamma^a \), satisfying \( \{ \Gamma^\mu, \Gamma^\nu \} = 2g^{\mu\nu} \), and $\overline{\psi} \equiv \psi^\dagger \gamma^0$ denotes the Dirac adjoint spinor. The covariant derivative \( \nabla_\mu \psi \) includes the spin connection to ensure general covariance. The remaining matter fields, collectively denoted by \( \psi_m \), are assumed to couple minimally to the metric only.
%, and hence interact only gravitationally with $\phi$ and $\psi$.

The field equations governing the dynamics of the system are derived by varying the total action with respect to the metric, the fermionic (spinor) field and the scalar field, respectively. This yields the following set of coupled equations \cite{Benisty:2019pxb,Lepe:2011sz}: 
\begin{eqnarray}
\label{eq2.3a}
&& R_{\mu\nu}-\frac{1}{2}g_{\mu\nu}R = T^{GB}_{\mu\nu}+ T^\phi_{\mu\nu} + T^{\psi}_{\mu\nu} + T_{\mu \nu}^{(\phi,\psi)} + 
T_{\mu\nu},\\
&& \nabla^2\phi= \frac{dV}{d\phi}-\alpha \frac{d f}{d\phi}{\cal G} + \frac{dM(\phi)}{d\phi} \overline{\psi} \psi - \frac{\partial \mathcal{L}_{\text{m}}}{\partial \phi},\\
&& i \, \Gamma^{\mu} \nabla_{\mu}\psi-M(\phi)\psi - \frac{\partial W}{\partial \overline{\psi}} +  \frac{\partial \mathcal{L}_{\text{m}}}{\partial \overline{\psi}}=0,\\
&&  i (\nabla_{\mu} \overline{\psi})\Gamma^{\mu} + M(\phi) \overline{\psi}  + \frac{\partial W}{\partial \psi} -  \frac{\partial \mathcal{L}_{\text{m}}}{\partial \psi}=0,
\end{eqnarray}
where $\nabla^2 \equiv g^{\mu\nu}\nabla_{\mu}\nabla_{\nu}$, and the corresponding energy momentum tensors are 
\begin{eqnarray}
\lb{eq2.4}
T^{GB}_{\mu\nu}& = &2\left(\nabla_\mu \nabla_\nu f\right)R-2g_{\mu\nu}\left(\nabla_\rho \nabla^\rho f\right)R 
		-4\left(\nabla^\rho \nabla_\nu f\right)R_{\mu\rho} \nonumber\\ && -4\left(\nabla^\rho \nabla_\mu f\right)R_{\nu\rho}+4\left(\nabla^\rho \nabla_\rho f\right)R_{\mu\nu}
		\nonumber\\ && +4g_{\mu\nu}\left(\nabla^\rho \nabla^\sigma f\right)R_{\rho\sigma} -4\left(\nabla^\rho \nabla^\sigma f\right)R_{\mu\rho\nu\sigma}\ ,\nonumber \\
	T_{\mu \nu}^{\phi} &=& \nabla_{\mu}\phi \nabla_{\nu} \phi - g_{\mu \nu} \bigg[\frac{1}{2}\nabla^{\alpha}\phi\nabla_{\alpha}\phi + V(\phi)\bigg],\nb\\	
 T_{\mu \nu}^{\psi} &=& \frac{i}{4} \left[ (\nabla_{\nu} \overline{\psi}) \Gamma_{\mu} \psi  + (\nabla_{\mu} \overline{\psi})\Gamma_{\nu}\psi  - \overline{\psi}\Gamma_{\mu} \nabla_{\nu}\psi - \overline{\psi}\Gamma_{\nu} \nabla_{\mu}\psi \right] \nonumber\\ && 
 -\frac{i}{2} g_{\mu \nu} \Big[(\nabla_{\lambda} \overline{\psi})\Gamma^{\lambda}\,\psi -\overline{\psi} \, \Gamma^{\lambda}\,\nabla_{\lambda} \psi \Big]  - g_{\mu \nu} W(\overline{\psi},\psi), \nb\\
 T_{\mu \nu}^{(\phi,\psi)} &=&- g_{\mu \nu} M(\phi) \overline{\psi} \psi,\nb \\
 T_{\mu\nu} & = & - \frac{2}{\sqrt{-g}}\frac{\delta \mathcal{L}_{\text{m}}}{\delta g^{\mu\nu}}.
\end{eqnarray}
%Here, $u^\mu = (1,\, \vec{0})$ denotes the four-velocity of the fluid in the comoving frame, satisfying the condition $u^\mu u_\mu = -1$. $\rho_m$ and $ p_m$ represent the energy density and pressure of the background fluid, respectively.\\

\subsection{Homogeneous and Isotropic Background}

In the following analysis, we adopt a spatially flat Friedmann–Lemaître–Robertson–Walker (FLRW) universe, described by the line element
\begin{equation}
    ds^2 = dt^2 - a(t)^2 \left( dx^2 + dy^2 + dz^2 \right),
\end{equation}
where \( a(t) \) is the cosmic scale factor, and the Gauss–Bonnet invariant simplifies to:
\begin{equation}
    \mathcal{G} = 24H^2(\dot{H} + H^2),
\end{equation}
where \( H = \dot{a}/a \) denotes the Hubble parameter. The other matter fields are assumed to be described by
\bqn
\lb{eq2.9}
T_{\mu\nu} & = & \rho_m\, u_\mu u_\nu + p_m\, (g_{\mu\nu} + u_\mu u_\nu),
\end{eqnarray}
where $u^\mu = (1,\, \vec{0})$ denotes the four-velocity of the fluid in the comoving frame, satisfying the condition $u^\mu u_\mu = -1$. $\rho_m$ and $ p_m$ represent the energy density and pressure of the background fluid, respectively.

Then,  the field equations reduce to the following set of coupled differential equations:
\begin{eqnarray}
&& \ddot{\phi}  + 3 H \dot{\phi} + \frac{d V(\phi)}{d \phi}  - \frac{df(\phi)}{d\phi} G + \frac{dM(\phi)}{d\phi} \overline{\psi} \psi = 0,\\
&& \dot{\psi}  + \frac{3}{2} H \psi + i \gamma^{0} \frac{\partial W}{\partial \overline{\psi}}  + i M(\phi) \gamma^{0}  \psi = 0,\\
&& \dot{\overline{\psi}}  + \frac{3}{2} H \overline{\psi} - i \frac{\partial W}{\partial \psi}  - i  M(\phi)  \overline{\psi}\gamma^{0} = 0.
 \end{eqnarray}
The corresponding   energy densities of the scalar and fermionic fields are defined by:
\begin{eqnarray}
 \rho_{\phi} &=& \frac{1}{2} \dot{\phi}^2 + V(\phi),\\
  p_{\phi}  &=& \frac{1}{2} \dot{\phi}^2 - V(\phi),\\
   \label{rhochi}
    \rho_{\psi}  &=& W(\psi, \overline{\psi}) + M(\phi) \psi \overline{\psi},\\
    p_{\psi}  &=&  \frac{\partial W}{\partial \psi} \frac{\psi}{2} + \frac{\partial W}{\partial \overline{\psi}} \frac{\overline{\psi}}{2}- W(\psi, \overline{\psi}).
\end{eqnarray}
The first Friedmann equation takes the form: 
\begin{equation}
	3 H^2 = \rho_{\phi} + \rho_{\psi} - 24 \dot{f}(\phi) H^3\, + \rho_{m}, 
\end{equation}
where $\dot{f}(\phi)$ represents the derivative of $f$ with respect to cosmic time $t$ and $\rho_m$ is the energy density of pressureless baryonic matter. The evolution equations for the energy densities of the scalar and fermionic fields, the 
baryonic matter and radiation are
\begin{eqnarray}
\label{rhophi}
    \dot{\rho}_{\phi} +  3H(\rho_{\phi}+p_{\phi}) &=& f'(\phi) \dot{\phi}G -Q\, ,  \\
    \dot{\rho}_{\psi} +  3H(\rho_{\psi}+p_{\psi}) &=& Q \, ,\\
     \dot{\rho}_{m} + 3 H \rho_{m}&=&0 \,, \\
    \dot{\rho}_{r} + 4 H \rho_{r}&=&0 \, ,
\end{eqnarray}
where $f'(\phi) \equiv df(\phi)/d\phi$ and \( Q \equiv M'(\phi) \, \dot{\phi} \, \overline{\psi} \psi \) denotes the interaction term between the two dark components.
%the scalar and fermionic sectors. 

In this framework, the interaction results in a dynamical exchange of energy within the dark sector, 
%. This exchange is often characterized by an effective interaction term \( Q \), 
which modifies the standard conservation equations. The sign of \( Q \) determines the direction of energy transfer: \( Q > 0 \) corresponds to energy flowing from dark energy to dark matter, while \( Q < 0 \) implies the reverse~\cite{Bolotin:2013jpa}. 
From a theoretical perspective, couplings between scalar and fermionic fields are well motivated. In effective field theories arising from unified theories,  such as string or M-theory, scalar degrees of freedom (e.g., dilatons) generically couple to matter fields. After dimensional reduction to four dimensions, these interactions typically induce scalar-dependent fermion masses or Yukawa-type couplings. Consequently, the fermionic energy density acquires a dependence on the scalar field, leading to a non-trivial exchange of energy between the scalar and fermionic sectors. In interacting dark sector models, allowing energy transfer between dark energy and dark matter provides a mechanism to address issues such as the coincidence problem and can modify the late-time expansion history in a controlled manner. Therefore, the coupling considered here is not imposed ad hoc, but is motivated both by high-energy theoretical constructions and by cosmological phenomenology. For more details, see, for example, \cite{Wang:2016lxa,Wang:2024vmw,Samojeden:2010rs,Ribas:2016ulz,Benisty:2019pxb,Grams:2014woa,Farrar:2003uw,Bean:2008ac,Feng:2010gw}, and references therein.

% Such couplings naturally happen in string/M-theory and impose constraints on possible forms of interactions, including those chosen at the phenomenological level.

%motivated by the quest for a more unified and fundamental description of the dark sector, beyond phenomenological parametrizations.

Equation~\eqref{rhophi} can be reformulated to include an effective pressure term, yielding:
\begin{eqnarray}
    \dot{\rho_{\phi}} + 3H(\rho_{\phi}+p^{\phi}_{\text{eff}}) = -Q \, ,
\end{eqnarray}
where the effective pressure is defined as:
\begin{eqnarray}
    p^{\phi}_{\text{eff}} \equiv p_{\phi} - 8H\dot{\phi}(\dot{H}+H^2)f'(\phi) \,.
\end{eqnarray}
Based on these definitions, the equations of state (EoS) parameters for the scalar and fermionic fields are given by:
\begin{eqnarray}
    w_{\phi} &=& \frac{p^{\phi}_{\text{eff}}}{\rho_{\phi}} = \frac{ \frac{1}{2} \dot{\phi}^2 - V(\phi)- 8H\dot{\phi}(\dot{H}+H^2)f'(\phi)}{\frac{1}{2} \dot{\phi}^2 + V(\phi)},~~~~~~\\
    w_{\psi} &=& \frac{p_{\psi}}{\rho_{\psi}} = \frac{ \frac{\psi}{2} \frac{\partial W}{\partial \psi}  + \frac{\overline{\psi}}{2}\frac{\partial W}{\partial \overline{\psi}} - W(\psi, \overline{\psi})}{W(\psi, \overline{\psi}) + M(\phi) \psi \overline{\psi}},\\
    w_{\text{tot}} &=& -1-\frac{2\dot{H}}{3H^2}.
\end{eqnarray}

The equation of state (EoS) parameter quantifies the ratio of pressure to energy density for a cosmological component and plays a central role in determining the Universe’s expansion dynamics. The total EoS parameter, \( w_{\text{tot}} \), reflects the combined effect of all fields on cosmic evolution. An accelerating Universe requires \( -1 \leq w_{tot} < -\tfrac{1}{3} \), while \( w_{tot} \simeq 0 \) characterizes a matter-dominated era. Values \( w_{tot} < -1 \) correspond to the phantom regime.

\subsection{Fermionic Field as  Dark Matter}

In the context of dark matter,  fermionic fields provide important and viable models. In particular, when coupled appropriately to a bosonic scalar field, a fermionic field can dynamically behave as a pressureless matter component \cite{Stiele:2010xz,Micheletti:2009jy,Lepe:2011sz}. The dynamical equations for the fermionic field $\psi$ and its adjoint $\overline{\psi}$, in the presence of a coupling to a bosonic scalar field, are given by (as discussed in the previous section):
\begin{eqnarray}
%\nonumber
\dot{\psi}  + \frac{3}{2} H \psi + i \gamma^{0} \frac{\partial W}{\partial \overline{\psi}}  + i M(\phi) \gamma^{0}  \psi &=& 0,\\
\dot{\overline{\psi}}  + \frac{3}{2} H \overline{\psi} - i \frac{\partial W}{\partial \psi} \gamma^{0}  - i  M(\phi) \overline{\psi} \gamma^{0} &=& 0,
\end{eqnarray}
where we assume that there is no direct interaction between $\psi$ and baryonic matter and radiation, except the gravitational ones. 

To investigate the energy evolution of the fermionic field, we analyze the time evolution of the bilinear quantity \(\psi \overline{\psi}\). Taking its time derivative and combining the fermionic field equations yields:
\begin{equation}
    \frac{d}{dt}(\psi \overline{\psi}) + 3H \psi \overline{\psi} = i \gamma^{0} \left( \frac{\partial W}{\partial \psi} \psi - \frac{\partial W}{\partial \overline{\psi}} \overline{\psi} \right).
\end{equation}

When the self-interaction potential takes the form \(W(\psi, \overline{\psi}) \sim (\psi \overline{\psi})^n\), the right-hand side vanishes due to the symmetry of the derivatives, resulting in:
\begin{equation}
    \frac{d}{dt}(\psi \overline{\psi}) + 3H \psi \overline{\psi} = 0\, .
\end{equation}
This immediately yields \( \psi \overline{\psi} \propto a^{-3} \). The scaling behavior of the total fermionic energy density given by Eq. \eqref{rhochi} depends on the weight between the self-interaction term $W(\psi, \bar{\psi})$ and the interaction (or the fermionic mass) term $M(\phi)\psi\bar{\psi}$. Since $\psi\,\overline{\psi} \propto a^{-3}$, the self-interaction term scales as $W \propto a^{-3n}$, while the mass term behaves as $M(\phi)\psi \overline{\psi} \propto \frac{M(\phi)}{a^{3}}$. 

In the regime where the self-interaction term dominates, the fermionic energy density scales as $\rho_\psi \propto a^{-3n}$,
with an effective equation-of-state parameter $w_\psi = n - 1$. 
Different choices of $n$ interpolate between distinct cosmological behaviors: $n=1$ corresponds to pressureless matter, 
$n>1$ leads to stiffer fluids (e.g., $n=2$ implies $w_\psi = 1$), 
and $n<2/3$ yields dark-energy--like behavior with $w_\psi < -1/3$. However, if the Yukawa-type interaction term $M(\phi)\bar{\psi}\psi$ dominates, 
the fermionic component instead behaves as interacting dark matter with
\begin{equation}
\rho_\psi \propto \frac{M(\phi)}{a^{3}}.
\end{equation}
In this case, the standard cold dark matter scaling $\rho_\psi \propto a^{-3}$ 
is recovered when $M(\phi)$ is constant, whereas a time-dependent $M(\phi)$ 
induces a dynamical dark matter mass that evolves with the scalar field. This distinction clarifies that the scaling $\rho_\psi \propto a^{-3n}$  applies strictly in the self-interaction--dominated regime, while the mass-dominated regime naturally realizes interacting dark matter behavior.

% Then, the energy density and equation of state (EoS) parameter of the fermionic field scale as \( \rho_{\psi} \propto a^{-3n} \) and \(w_{\psi} = n - 1 \), respectively. Thus, different choices of \( n \)  interpolate between various cosmological behaviors. In particular, \( n = 1 \) produces the scaling behavior of non-relativistic pressureless matter and reinforces the interpretation of the fermionic bilinear as an effective dark matter component. On the other hand, \( n > 1 \) yields stiffer fluids (e.g., when $n = 2$ we have \(w_{\psi} = 1 \)), while \( n < 2/3 \) leads to dark energy-like dynamics with \(w_{\psi} < -1/3 \).

% This framework not only provides a rich structure for modeling the dark sector but also offers a natural interpretation of fermionic fields as constituents of dark matter. By coupling a fermionic field \( \psi \) to a scalar sector—typically through a self-interaction potential \( W(\psi, \overline{\psi}) \) and a mass interaction term \( M(\phi) \), one introduces a mechanism by which dark matter can evolve dynamically over cosmic time. Notably, the interaction term renders the dark matter mass dependent on the quintessence field \( \phi \). As a result, the non-relativistic dark matter energy density does not redshift strictly as \( a^{-3} \), as in the case of ordinary matter, but instead evolves proportionally to \( M(\phi)/a^{3} \).

\section{Formulation of dynamical system}
\label{dynamics}

To investigate the dynamics of the non-minimally coupled system, we begin by introducing the following set of dimensionless dynamical variables:
\begin{eqnarray}
\nonumber
&&	x = \dfrac{\dot{\phi}}{\sqrt{6} H}, \quad y= \dfrac{\sqrt{V}}{\sqrt{3} H}, \quad z = \frac{\sqrt{W}}{\sqrt{3} H}, 
  u = \frac{\sqrt{M \psi \overline{\psi}}}{\sqrt{3} H}, \\&& \quad v = f_{,\phi} H^2,\quad \Omega_r = \frac{\rho_{r}}{3H^2} \quad \Omega_b = \frac{\rho_b}{3H^2}.
\end{eqnarray} 
In addition, we define the auxiliary variables:
\begin{eqnarray}
%\nonumber
&&	\Gamma_{1} = \frac{V_{,\phi}}{V},\quad \Gamma_{2} = \frac{M_{,\phi}}{M}, \quad  \Sigma_{1} = \frac{\psi \ W_{,\psi}}{W},   \nonumber \\
 &&    \Sigma_{2} = \frac{\overline{\psi} \ W_{, \overline{\psi}}}{W}, \quad A = \frac{f_{,\phi\phi}}{f_{,\phi}}, \quad B = \frac{i \gamma^{0} M}{H}.
\end{eqnarray}
With these definitions, the Friedmann constraint equation takes the form:
\begin{equation} \label{con}
	1 = x^2 + y^{2} + z^{2} + u^{2} + \Omega_r + \Omega_b - 8 \sqrt{6}\left( x v \right).
\end{equation}
The system can then be reformulated as an autonomous set of first-order differential equations:
\begin{eqnarray}
\label{dy1}
	x' & = & -3 x - \frac{3 \Gamma_{1} y^2}{\sqrt{6}} + \frac{24 v}{\sqrt{6}} - \frac{3 \Gamma_{2} u^2}{\sqrt{6}} + \frac{\dot{H}}{H^2} \left(\frac{24}{\sqrt{6}} v - x\right),~~~~~~\\
    \label{dy2}
	y' & = & \frac{\sqrt{6} \Gamma_{1} x y}{2}  - y \frac{\dot{H}}{H^2}\, ,\\
    \label{dy3}
	z' &= & -\frac{3 z}{4} \left(\Sigma_{1} + \Sigma_{2} \right)  + \frac{ z \ B}{2}\left(\Sigma_{2} - \Sigma_{1} \right) - z \frac{\dot{H}}{H^2},\\
    \label{dy4}
	u' & =& \frac{\sqrt{6}}{2} \Gamma_{2} x u - \frac{3}{2} u + \frac{1}{2} z^2 \frac{B}{u} \left( \Sigma_{1} -\Sigma_{2} \right) -   u \frac{\dot{H}}{H^2}\, ,\\
    \label{dy5}
	v' & =& \sqrt{6} A x v + 2v \frac{\dot{H}}{H^2}\ ,\\
    \label{dy6}
        B' &=& \sqrt{6} B \Gamma_{2}x - B \frac{\dot{H}}{H^2},\\
        \label{dy7}
        \Omega_r' &=& -4\Omega_r - 2\Omega_r \frac{\dot{H}}{H^2},\\
        \label{dy8}
        \Omega_b' &=& -3\Omega_b - 2\Omega_b \frac{\dot{H}}{H^2}.
\end{eqnarray}
We denote a derivative with respect to \( N = \ln a \) as \( ()' = \frac{d()}{dN} = \frac{1}{H} \frac{d()}{dt} \), and note that \( dN = H\,dt = d\ln a \).

Additionally, the tensor propagation speed \( c_T^2 \), which depends on the Gauss–Bonnet coupling function \( f(\phi) \), is given by \cite{Tsujikawa:2006ph, Odintsov:2019clh,Fier:2025huc}:
\begin{equation}
    c_T^2 = \frac{1 + 8\kappa^2 \ddot{f}}{1 + 8\kappa^2 H \dot{f}} \,.
\end{equation}
Recent gravitational wave observations have confirmed that the propagation speed of gravitational waves closely matches the speed of light \cite{LIGOScientific:2017zic}, placing the stringent constraint:
\begin{equation}
\label{ct}
    |c_T^2 - 1| \leq 5 \times 10^{-16},
\end{equation}
in natural units where \( c = 1 \) \cite{TerenteDiaz:2023iqk}. Furthermore, to avoid ghost instabilities at the perturbative level, \( c_T^2 \) must remain positive.

% To incorporate these physical constraints and simplify the mathematical complexity of the system, we impose the condition:
% \begin{equation}
%     c_T^2 = 1 \quad \Rightarrow \quad \dot{f} \propto a.
% \end{equation}
% This assumption makes the framework effectively model-independent. Consequently, the time evolution of the Gauss–Bonnet coupling function is taken as:
% \begin{equation}
%     \dot{f}(\phi) = m_1 a, \quad \ddot{f} = m_1 a,
% \end{equation}
% where \( m_1 \) is a constant with mass dimension.

In the following analysis, we examine two distinct scenarios based on the choice of the potential \( V(\phi) \) and $M(\phi)$. In the first case, we explore the dynamics when \( c_T^2 \neq 1 \), while in the second, we consider the simplified case where \( c_T^2 = 1 \).  This comparative approach allows us to assess how the condition on tensor speed influences the evolution and stability of the cosmological model. Note that in the first case we still impose the condition (\ref{ct}), while in the second case this condition is satisfied identically.

\subsection{Model I}

In accordance with the previous analysis, we proceed to derive the autonomous system of equations for a specific choice of the potential and coupling functions. To enable a tractable and physically meaningful study, we adopt the following functional forms \cite{Skugoreva:2013ooa,Saridakis:2009pj,Ribas:2016ulz,Benisty:2019pxb}:
\begin{eqnarray}
\label{eq3.14}
&& V(\phi) = V_{0} \phi^2, \quad M(\phi) = \beta \phi, \quad \\
\label{eq3.15}
&& W = W_{0} \psi \overline{\psi}, \quad  f(\phi) = \alpha \phi,
\end{eqnarray}
where $V_0$ \footnote{For a general potential of the form $V(\phi) = k \phi^{\delta}$, the constant $k$ has units of $m^{4+\delta}$.}, $\beta$, $W_0$, and $\alpha$ are constant parameters that govern the strength and nature of the scalar potential and various couplings in the model. However, $\phi$ has a mass dimension of one. This makes $\alpha$ a dimension of [M]$^{-1}$.

These choices facilitate a detailed and consistent investigation of the system's dynamical behavior. Under these assumptions, the auxiliary variables simplify to:
\begin{equation}
%\nonumber
	\Gamma_{1} = \frac{2}{\phi},\   \Sigma_{1} = 1 = \ \Sigma_{2}, \ \Gamma_{2} = \frac{1}{\phi}, \ A = 0. 
\end{equation}
This leads to the relations:
\begin{equation}
    \Gamma_{1} = 2 \Gamma_{2} \quad \text{and} \quad \phi = \frac{W_{0} u^2}{\beta z^2}.
\end{equation}
Accordingly, the autonomous system takes the forms %is modified as follows:
\begin{eqnarray}
\label{dyMI1}
	x' & = & -3 x - \frac{3 \Gamma_{1} y^2}{\sqrt{6}} + \frac{24 v}{\sqrt{6}} - \frac{3 \Gamma_{2} u^2}{\sqrt{6}} + \frac{\dot{H}}{H^2} \left(\frac{24}{\sqrt{6}} v - x\right),\nb\\
    ~~~~~~~~~~~\\
    \label{dyMI2}
	y' & = & \frac{\sqrt{6} \Gamma_{1} x y}{2}  - y \frac{\dot{H}}{H^2}\, ,\\
    \label{dyMI3}
	z' &= & -\frac{3 z}{4} \left(\Sigma_{1} + \Sigma_{2} \right) - z \frac{\dot{H}}{H^2},\\
    \label{dyMI4}
	u' & =& \frac{\sqrt{6}}{2} \Gamma_{2} x u - \frac{3}{2} u -   u \frac{\dot{H}}{H^2}\, ,\\
    \label{dyMI5}
	v' & =&  2v \frac{\dot{H}}{H^2}\ ,\\
    \label{dyMI6}
     B' &=& \sqrt{6} B \Gamma_{2}x - B \frac{\dot{H}}{H^2}\, ,\\
     \label{dyMI7}
       \Omega_r' &=& -4\Omega_r - 2\Omega_r \frac{\dot{H}}{H^2},\\
       \label{dyMI8}
        \Omega_b' &=& -3\Omega_b - 2\Omega_b \frac{\dot{H}}{H^2}.
\end{eqnarray}
and the Hubble derivative is given by:
\bqn
&&\frac{\dot{H}}{H^2} = \frac{1}{2 \left(1 + 8\sqrt{6}\, x v + 96\, v^2\right)} \Big[ 
32 \sqrt{6} x v - 6 x^2  -3u^2 
- 4 \Omega_r  \nonumber  \\
&& -3 \Omega_b- 192 v^2  + 24 \Gamma_1 v y^2 + 24 \Gamma_2 v u^2 - \frac{3}{2}\, z^2 (\Sigma_1 + \Sigma_2)\Big].\nb\\
\eqn
The fractional energy densities of the scalar field, fermionic field, and radiation are defined as:
\begin{eqnarray}
    \Omega_{\phi} = x^2+y^2, \quad \Omega_{\psi} = z^2+u^2.
\end{eqnarray}
and they satisfy the following constraint:
\begin{eqnarray} \label{con1}
    \Omega_{\phi}+\Omega_{\psi} + \Omega_{r} + \Omega_b-8\sqrt{6}xv = 1.
\end{eqnarray}
It is essential to recognize that not all defined variables are independent. Some can be expressed in terms of others, with only the truly independent variables, those whose evolution must be determined, treated as primary.
From the structure of the autonomous equations, it is evident that the primary variables driving the dynamics of the system are $\left(x,y,z,v,B,\varrho\right)$. One can observe that the variable $u$ can be expressed in terms of other variables using equation \eqref{con}.\\

To solve the system, it is crucial to account for the constraints imposed on the dynamical variables. In particular, the energy densities must remain positive and bounded throughout the evolution, namely:
\begin{eqnarray}
\label{density}
   0 \leq \Omega_{\phi} \leq 1, \quad  0 \leq \Omega_{\psi} \leq 1, \quad 0 \leq \Omega_{r} \leq 1,
\end{eqnarray}
which means $0 \leq x^2+y^2 \leq 1$ and $0 \leq z^2 + u^2 \leq 1$.
The bounds in Eq.~\eqref{density} define the physically admissible region of the phase space, corresponding to non-negative and finite energy densities consistent with the Friedmann constraint~\eqref{con1}. These conditions are not imposed as an artificial truncation of the phase space. Rather, they follow directly from the normalization of the total energy density with respect to the critical density. In the dynamical analysis, the system evolves on the constraint hypersurface determined by Eq.~\eqref{con1}. The autonomous equations preserve this constraint under time evolution, ensuring that physically viable trajectories remain within the region $0 \le \Omega_i \le 1$. Hence, the phase space is naturally restricted to the physically meaningful subset defined by the Friedmann relation, and the bounds reflect dynamical consistency rather than an externally imposed cut. In Appendix A, we provide a detailed numerical analysis of the autonomous system.

\begin{figure*}[htbp]
\centering
\includegraphics[scale = 0.35]{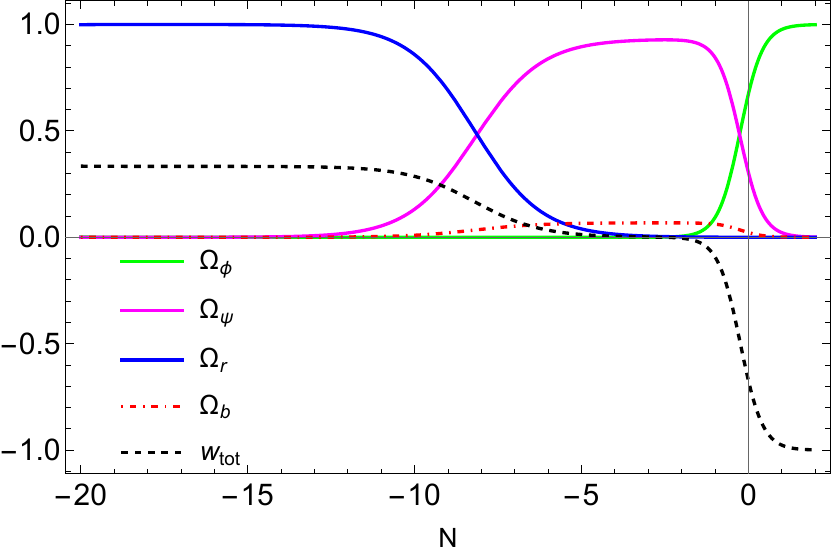} \hspace{0.1in}
\includegraphics[scale = 0.38]{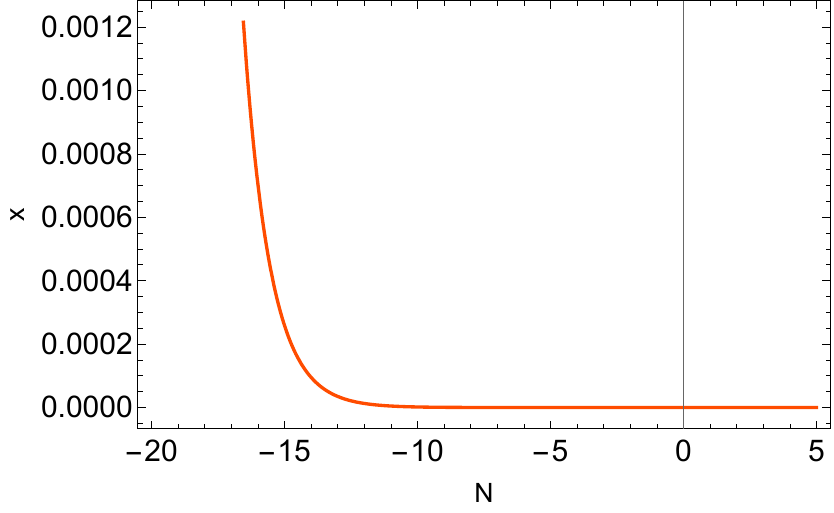}  \hspace{0.1in}
\includegraphics[scale = 0.36]{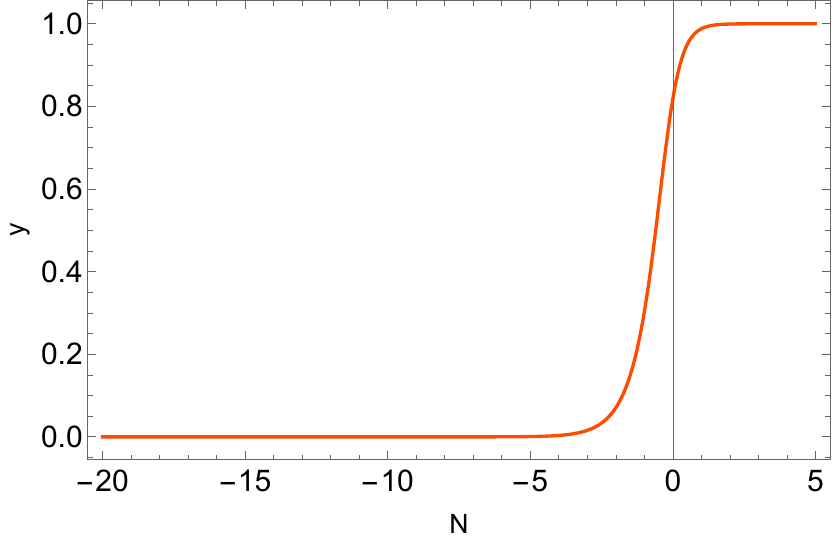} \hspace{0.1in}
\includegraphics[scale = 0.35]{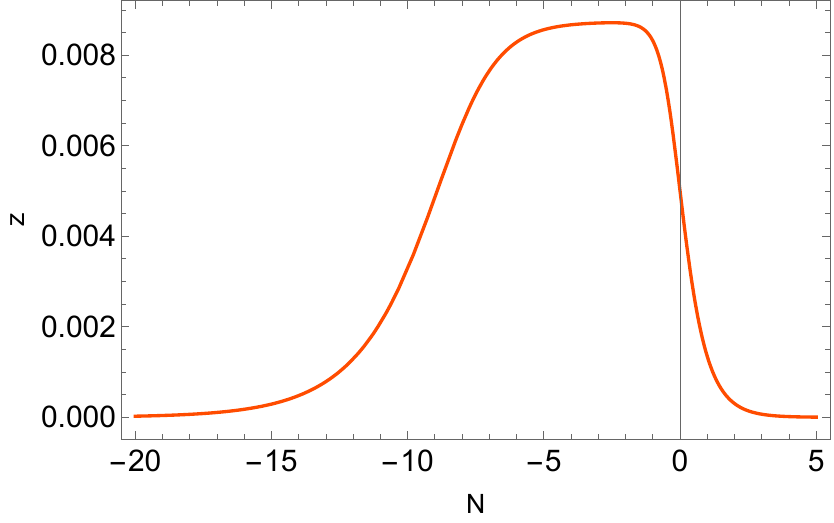} \hspace{0.1in}
\includegraphics[scale = 0.35]{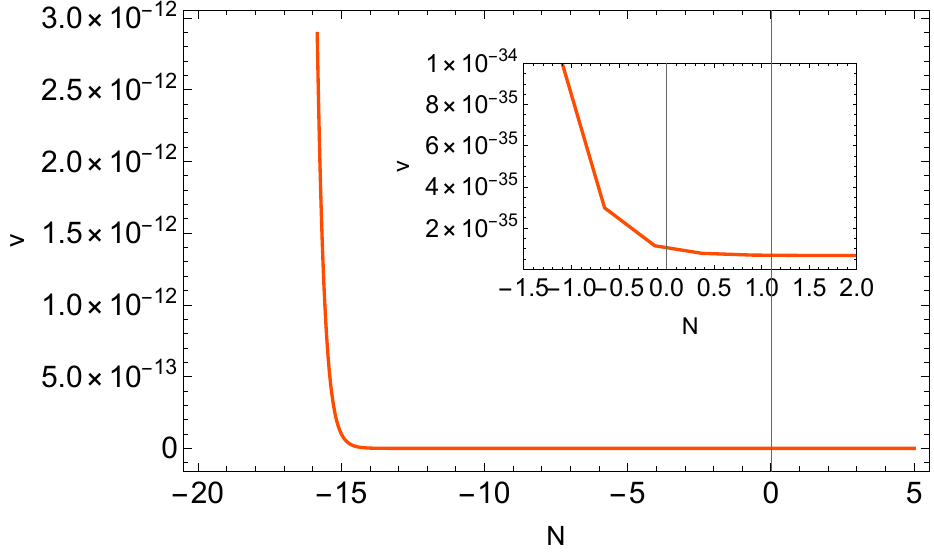} \hspace{0.1in}
\includegraphics[scale = 0.38]{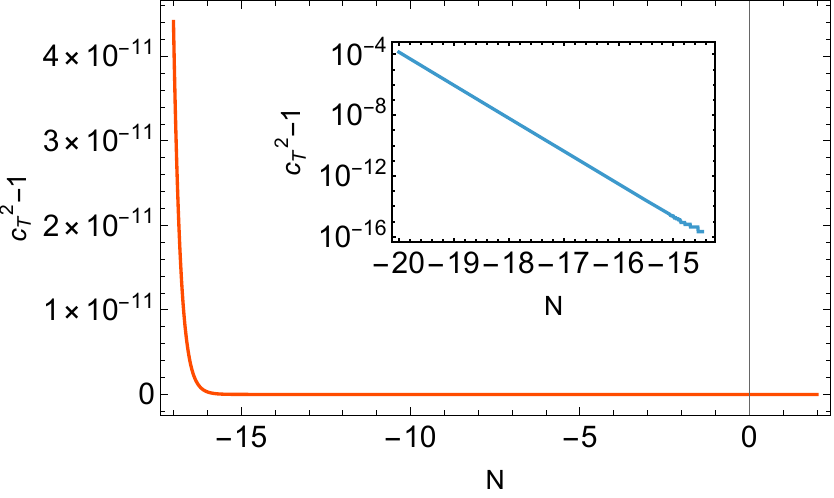}
\caption{The evolution of the cosmological parameters \( \Omega_{\phi} \), \( \Omega_{\psi} \), \( \Omega_{r} \), $\Omega_b$ and \( w_{\text{tot}} \) for Model I is shown as a function of \( N \). The system is initialized with the parameter values: \( \beta = 10^{-8} \), \( W_0 = 1 \), \( x_0 = 10^{-13} \), \( y_0 = 0.82 \), \( z_0 = 0.005 \), \( v_0 = 10^{-35} \), $B=0.1$, \( \Omega_{r,0} = 0.0009 \), $\Omega_{b,0}h^2=0.0223$. We also examine the evolution of the gravitational wave (GW) constraint for the same initial configuration, ensuring consistency with observational limits. The main panel shows the full cosmological evolution up to the present epoch, where $c_{T}^2 \rightarrow 1$ The inset displays the early-time behavior on a logarithmic scale, clearly demonstrating a smooth exponential decay. The slight irregularity near $10^{-16}$ arises from the machine-precision limit of the numerical solver and is not physical.}
\label{ModelA}
\end{figure*}

We require the model to reproduce a realistic cosmological history, namely a radiation-dominated era followed by matter domination and a late-time dark-energy phase, while yielding present-day density parameters $\Omega_{n0}$ ($n=r,\psi,\phi$) consistent with observational constraints. Radiation domination must extend at least to the epoch of Big Bang
Nucleosynthesis (BBN), which occurs at 
$z_{\mathrm{BBN}} \approx 4 \times 10^{8}$, corresponding approximately to $N \simeq -20$. Imposing this requirement significantly restricts the allowed phase-space trajectories. In particular, once a genuine radiation era is realized, a subsequent matter-dominated epoch with amplitude and duration close to that of $\Lambda$CDM emerges automatically. Moreover, viable radiation-era solutions constrain the scalar-field equation-of-state parameter $w_\phi$, thereby reducing
the degree of fine tuning. Radiation-matter equality, defined by $\Omega_r = \Omega_\psi$, occurs at
\begin{equation}
N_{r=m} = \ln\left(\frac{\Omega_{r0}}{\Omega_{\psi0}}\right)
\approx -8.1,
\end{equation}
consistent with the standard cosmological timeline. At later times, when radiation becomes negligible, matter-dark energy
equality ($\Omega_\psi = \Omega_\phi$) is obtained at
\begin{equation}
N_{\phi=\psi}
= \frac{1}{3}
\ln\left(
\frac{\Omega_{\psi0}}{\Omega_{\phi0}}
\frac{\rho_{\phi0}}{\rho_\phi}
\right),
\end{equation}
where $\rho_\phi$ is evaluated at equality. Numerically, we find
$N_{\phi=\psi} \approx -0.2$. We now turn to the dynamical evolution shown in Figure~\ref{ModelA}. The figure displays the fractional energy densities (\( \Omega_{\phi} \)), fermionic matter (\( \Omega_{\psi} \)), and radiation (\( \Omega_{r} \)) as function of $N=\text{ln}\,a$, together with the effective total equation-of-state parameter. This combined representation allows a consistent assessment of the model across all cosmological epochs. From the numerical evolution, the model successfully reproduces the standard sequence of radiation-, matter-, and dark-energy–dominated eras in quantitative agreement with observational constraints. At early times ($N \simeq -20$), the Universe is radiation dominated with $\Omega_r \approx 1$ and $\Omega_\phi \ll 10^{-4}$, ensuring that the scalar-field contribution is negligible during the Big Bang Nucleosynthesis (BBN) epoch and therefore does not affect the standard light-element abundance predictions. Radiation–matter equality occurs at $N_{\rm eq} \simeq -8.2$, corresponding to $z_{\rm eq} \simeq 3600$, which is consistent with the Planck 2018 estimate. The matter-dominated phase extends approximately from $N \simeq -8.2$ to $N \simeq -1.2$, yielding a duration $\Delta N \approx 7$, sufficient to support structure formation. During this interval, the effective equation of state approaches $w_{\rm tot} \approx 0$, as expected for a pressureless matter era. The present baryon fraction remains consistent with observational bounds, $\Omega_{b0} \sim 0.02$, ensuring a realistic matter composition. The onset of cosmic acceleration occurs at $N_{\rm acc} \simeq -0.6$, where $w_{\rm tot}$ crosses the threshold $-1/3$, in agreement with observational estimates. At late times, the system evolves toward a stable accelerated attractor characterized by 
$\Omega_\phi \to 1$ and $w_{\rm tot} \to -1$.
We find that the model yields observationally viable behavior when the following conditions are satisfied: \( v \leq 10^{-23} \) for the Gauss–Bonnet coupling term. The upper bound on the GB coupling is further supported by the evolution of \( c_{T}^2 \), which remains within the constraint defined in equation~\eqref{ct}. The inset shows the same quantity on a logarithmic scale, which makes the exponential suppression at early times manifest. In our numerical solution, we explicitly verify that the bound remains satisfied at $a=1$, corresponding to the redshift range relevant for gravitational-wave observations.

An important feature of our model is that the GB coupling function dynamically evolves toward very small values at low redshifts once observational constraints are imposed. Consequently, the GB contribution becomes negligible at late times, effectively recovering GR at the present epoch. At higher redshifts, however, the coupling strength increases, allowing for non-trivial modifications of the cosmological dynamics. Nevertheless, the requirement of ghost and gradient stability restricts the allowed parameter space such that the GB term remains sufficiently controlled and does not drive the effective equation of state into the phantom regime. Throughout the cosmic evolution, the interacting models closely mimic $\Lambda$CDM at the background level, with the scalar-field equation of state remaining in the accelerating regime and free from ghost instabilities. Thus, the late-time suppression of the GB term reflects a balance between theoretical consistency and observational viability: the modification is dynamically relevant in the asymptotic past, yet naturally suppressed at low redshifts to recover GR.

In Appendix A, we consider the robustness of the model of Fig. \ref{ModelA} by varying the initial conditions, and find that with any of these initial conditions  a viable model exists, that is, the standard  cosmological sequence of radiation domination, followed by matter domination and late-time acceleration, is always preserved. For more details, see the analysis provided in Appendix A.

Incorporating interactions within the dark sector significantly improves the Universe’s evolutionary history, not only in the late-time but also during the radiation- and matter-dominated eras. Moreover, since the interaction mediates the exchange between dark matter and dark energy, it should be constrained so that, in the asymptotic past, it does not alter the characteristic profile of the radiation phase. This prolonged matter-dominated phase is crucial for the large-scale structure formation.

\begin{figure}[htbp]
    \centering
        \includegraphics[scale = 0.45]{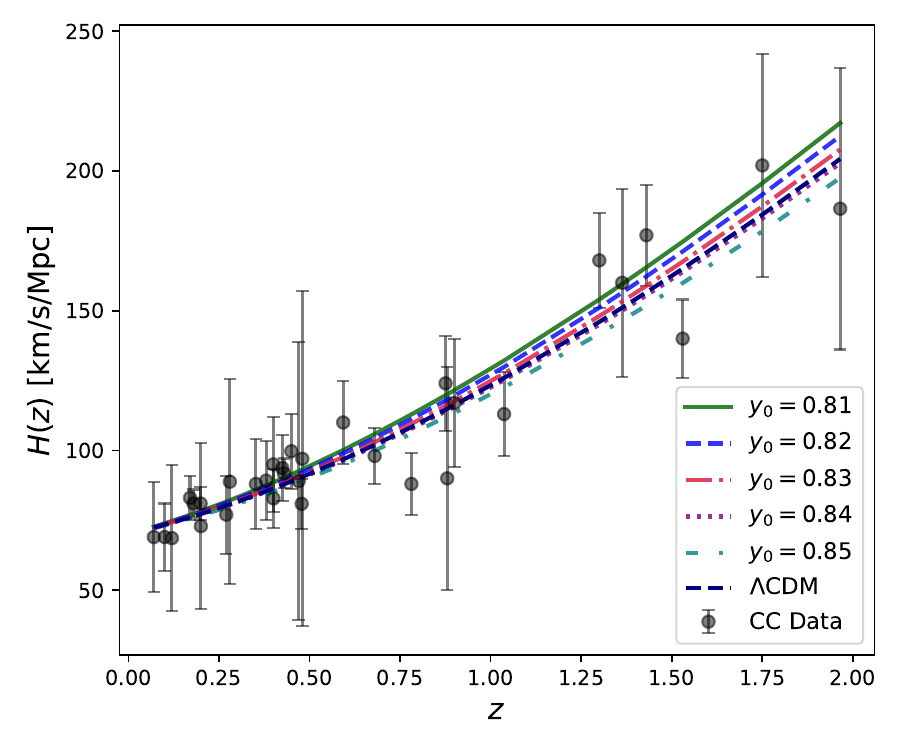} \hspace{0.1cm}
    \caption{Evolution of the Hubble parameter \( H(z) \) as a function of redshift \( z \) for distinct initial conditions \( y_0 \) for Model I. The curves correspond to theoretical predictions for \( y_0 \in (0.81, 0.85) \), along with the standard \( \Lambda \)CDM model. The data points with error bars represent cosmic chronometer (CC) measurements of \( H(z) \). The plot illustrates the impact of varying \( y_0 \) on the expansion history and its consistency with observational data. All model curves are generated for the fixed value of \( H_0 = 70\, \text{km/s/Mpc} \).
}
    \label{ModelHA}
\end{figure}

Figure~\ref{ModelHA} presents a quantitative comparison between the expansion history predicted by the modified model and the standard \( \Lambda \)CDM scenario. The observational data points with error bars show good agreement with the theoretical predictions. The colored curves represent the Hubble parameter \( H(z) \) for different values for the initial condition \( y_0 \equiv y(N=0)\), varied between 0.81 and 0.85 due to its sensitivity to the expansion rate. The close clustering of the curves indicates that variations in \( y_0 \) within this range lead to only mild changes in the expansion history. At low redshifts, deviations from \( \Lambda \)CDM remain controlled, demonstrating the model’s consistency with current observations while allowing for modified dynamics at earlier times.

\subsection{Model II}

In contrast to the previous subsection, where no assumption was made about the tensor propagation speed \( c_{T}^2 \), we now proceed under the constraint \( c_{T}^2 = 1 \) to comply with observational bounds from gravitational wave events. This assumption simplifies the analysis and enables a more model-independent formulation. As shown earlier, the tensor speed is given by:
\begin{eqnarray}
    \ct &= &  \frac{1 + 8 \kappa^2 \ddot{f}}{1 + 8 \kappa^2 H \dot{f}} = 1.
\end{eqnarray}
Imposing this condition leads to a specific constraint on the Gauss–Bonnet coupling function \( f(\phi) \):
\begin{equation}
    \dot{f}(\phi) = \alpha\, a, \quad \ddot{f} = \alpha\, \dot{a},
\end{equation}
where \( \alpha \) is an integration constant with the mass dimension. Since \( \dot{f} \) scales with the scale factor and \( f = f(\phi) \), the derivatives of \( f \) with respect to the scalar field \( \phi \) can be expressed as:
\begin{eqnarray}
    f_{,\phi} = \frac{\alpha \ a}{\dot{\phi}}, \quad f_{,\phi \phi}= \frac{\alpha}{\dot{\phi}^2}\left(\ \dot{a} - \ a\ \frac{\ddot{\phi}}{\dot{\phi}}   \right).
\end{eqnarray}

We consider a class of models in which the potentials take the following forms, motivated by previous studies~\cite{Pourtsidou:2013nha,Ribas:2010zj,Pavluchenko:2003ge}: 
\begin{equation}
V(\phi) = \tilde{V}_{0} \,e^{-\gamma \phi},  \quad W(\psi,\overline{\psi}) = \tilde{\lambda} \psi \overline{\psi}, \quad M(\phi) = M_{0}  e^{\mu \phi},
\end{equation}
where, $\tilde{V_{0}}$, $\gamma$, $\tilde{\lambda}$, $M_{0}$, $\mu$ are free parameters. We further introduce a dimensionless parameter defined by
\begin{eqnarray}
    \Theta = \frac{\alpha a H}{\sqrt{1+\alpha^2 a^2 H^2}}.
\end{eqnarray}
This form of \( \Theta \) is chosen for both physical and mathematical reasons. It ensures that \( \Theta \in [0,1) \), thereby avoiding divergences in the dynamical system and maintaining a bounded phase space. Moreover, the expression remains well-behaved across different cosmological epochs \footnote{In the early universe (\( \alpha a H \ll 1 \)), \( \Theta \approx \alpha a H \), recovering the original scaling form \( \dot{f} = \alpha a \); in the late-time regime (\( \alpha a H \gg 1 \)), \( \Theta \to 1 \), effectively capturing the saturation of the Gauss-Bonnet contribution.}.

% \begin{equation} \label{con}
% 	1 = x^2 + y^{2} + z^{2} + u^{2} + \Omega_r + \Omega_b - 8 \sqrt{6}\left( x v \right).
% \end{equation}
% The system can then be reformulated as an autonomous set of first-order differential equations:
% \begin{eqnarray}
% \label{dy1}
% 	x' & = & -3 x - \frac{3 \Gamma_{1} y^2}{\sqrt{6}} + \frac{24 v}{\sqrt{6}} - \frac{3 \Gamma_{2} u^2}{\sqrt{6}} + \frac{\dot{H}}{H^2} \left(\frac{24}{\sqrt{6}} v - x\right),~~~~~~\\
%     \label{dy2}
% 	y' & = & \frac{\sqrt{6} \Gamma_{1} x y}{2}  - y \frac{\dot{H}}{H^2}\, ,\\
%     \label{dy3}
% 	z' &= & -\frac{3 z}{4} \left(\Sigma_{1} + \Sigma_{2} \right)  + \frac{ z \ B}{2}\left(\Sigma_{2} - \Sigma_{1} \right) - z \frac{\dot{H}}{H^2},\\
%     \label{dy4}
% 	u' & =& \frac{\sqrt{6}}{2} \Gamma_{2} x u - \frac{3}{2} u + \frac{1}{2} z^2 \frac{B}{u} \left( \Sigma_{1} -\Sigma_{2} \right) -   u \frac{\dot{H}}{H^2}\, ,\\
%     \label{dy5}
% 	v' & =& \sqrt{6} A x v + 2v \frac{\dot{H}}{H^2}\ ,\\
%     \label{dy6}
%         B' &=& \sqrt{6} B \Gamma_{2}x - B \frac{\dot{H}}{H^2},\\
%         \label{dy7}
%         \Omega_r' &=& -4\Omega_r - 2\Omega_r \frac{\dot{H}}{H^2},\\
%         \label{dy8}
%         \Omega_b' &=& -3\Omega_b - 2\Omega_b \frac{\dot{H}}{H^2}.
% \end{eqnarray}

Next, we define a new set of variables to facilitate the construction of the autonomous system:
\begin{equation}
	\Gamma_{1} = -\gamma,\ \   \Sigma_{1} = 1 = \Sigma_{2}, \ \ \Gamma_{2} = \mu, \ \  v =  \frac{\Theta}{x \sqrt{6}\sqrt{1 - \Theta^2}}.
\end{equation}
Hence, the only modified equation in the autonomous system, given by Eqs.(\ref{con}) - (\ref{dy8}), is the replacement of Eq.(\ref{dy5}) by
\begin{eqnarray}
    \Theta' & =& \Theta (1 - \Theta^2) \left(1 + \frac{\dot{H}}{H^2}\right) \, ,
\end{eqnarray}
 while the derivative of the Hubble parameter is given by:
\begin{eqnarray}
\nonumber
    \frac{\dot{H}}{H^2} = -\frac{\left[ 6 x^2 + (3/2)z^2(\Sigma_{1}+\Sigma_{2})+3u^2 + 4 \Omega_r + 3 \Omega_b\right]}{2\left(1+\frac{8\Theta}{\sqrt{1-\Theta^2}}\right)}.
\end{eqnarray}

\begin{figure*}
    \centering
    \includegraphics[scale = 0.32]{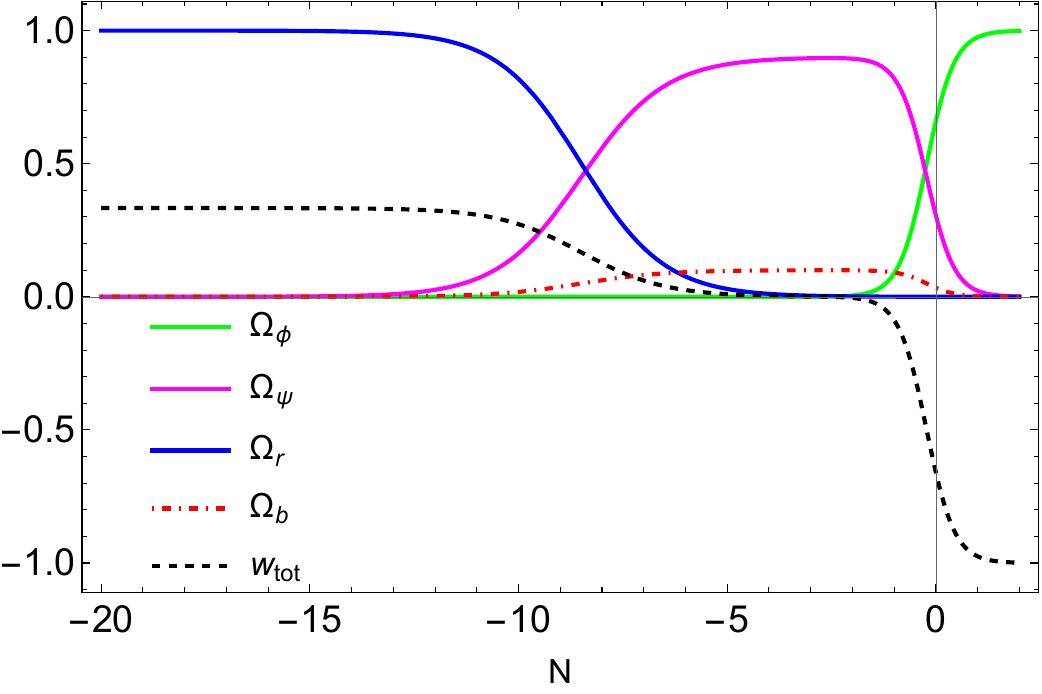} \hspace{0.1cm}    \includegraphics[scale = 0.33]{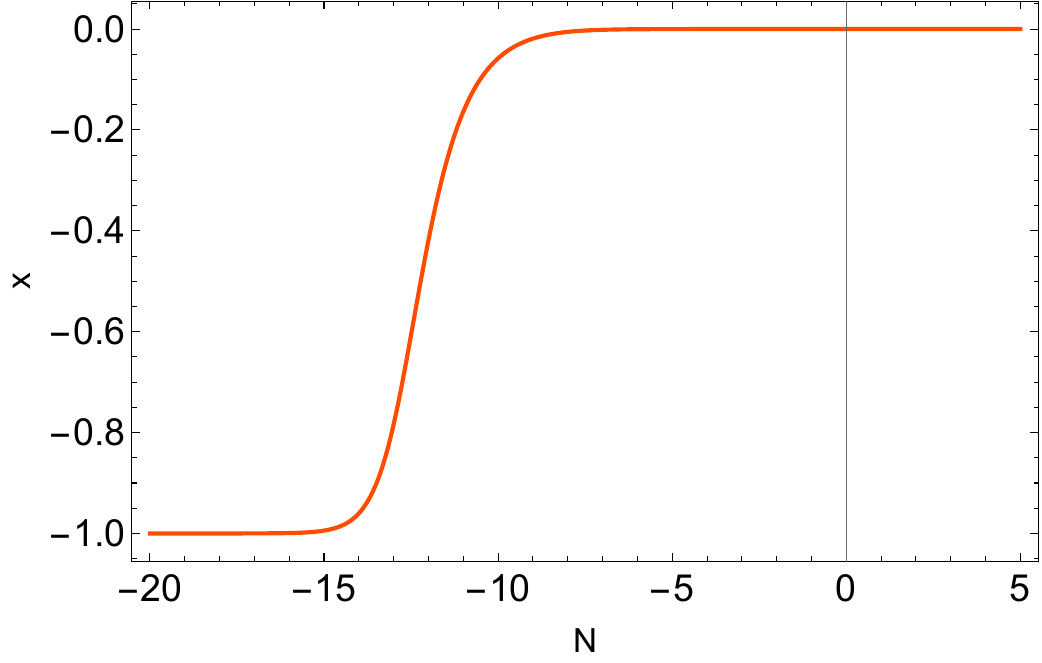}  \hspace{0.1in}
\includegraphics[scale = 0.32]{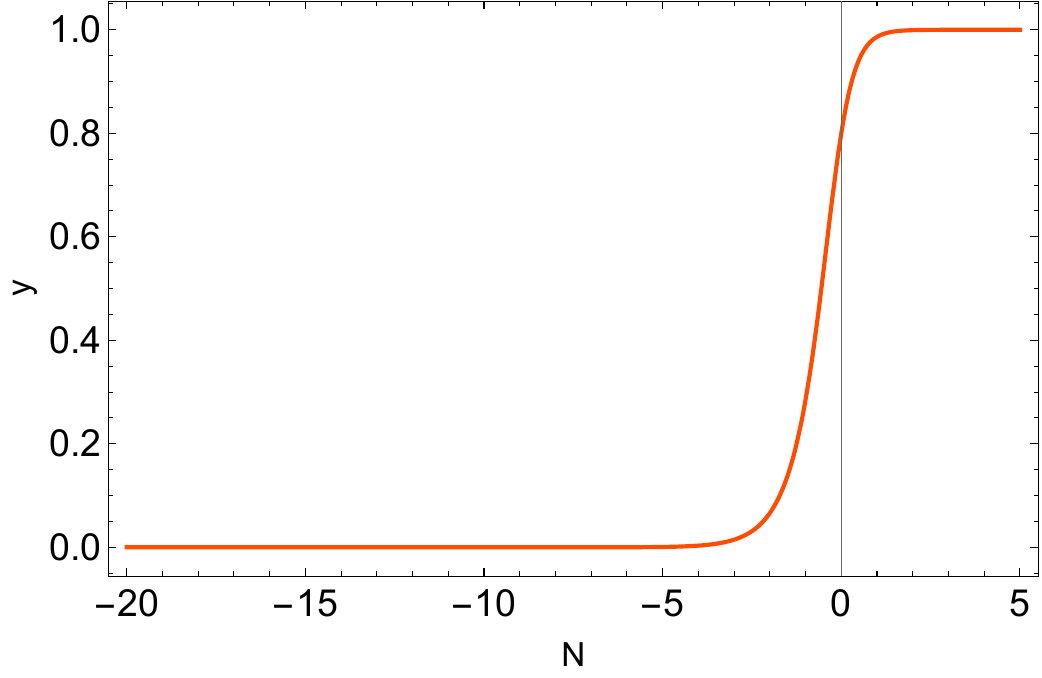} \hspace{0.1in}
\includegraphics[scale = 0.3]{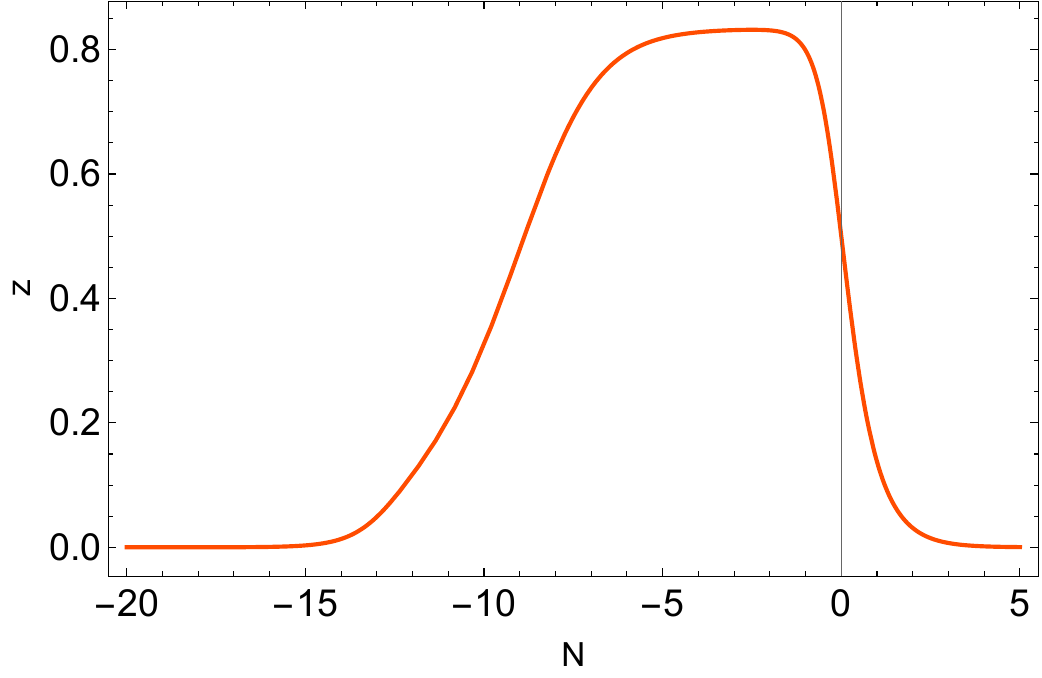} \hspace{0.1in}
\includegraphics[scale = 0.33]{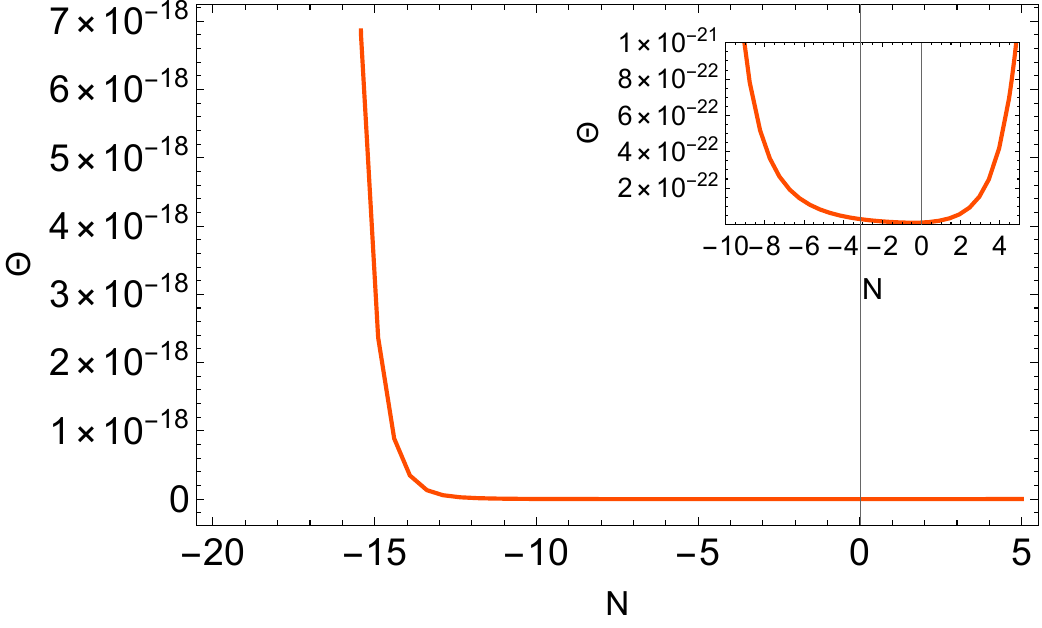} \hspace{0.1in}
    \caption{The evolution of the cosmological parameters \( \Omega_{\phi} \), \( \Omega_{\psi} \), \( \Omega_{r} \), $\Omega_b$ and \( w_{\text{tot}} \) for Model II is shown as a function of \( N \) in which we set $c_T = 1$, so the gravitational wave constraint is identically satisfied. In order to explicitly realize a radiation-dominated epoch and verify consistency with the standard cosmological sequence, the density-parameter evolution is obtained by imposing initial conditions deep in the radiation era at $N=-20$ and evolving the system forward. For completeness, the dynamical phase-space variables are displayed and the system is initialized at the present epoch with parameter values: $\gamma=10^{-6}$, $\mu=- 10^{-4}$, $x_0=10^{-5}$, $y_0 = 0.8$, $z_0 = 0.5$, $\Theta_0 = 10^{-23}$, $\Omega_{r0}  = 0.00009$, $\Omega_{b0}=0.223$. }
    \label{fig:ModelB}
\end{figure*}

\begin{figure}[htbp]
    \centering
    \includegraphics[scale = 0.45]{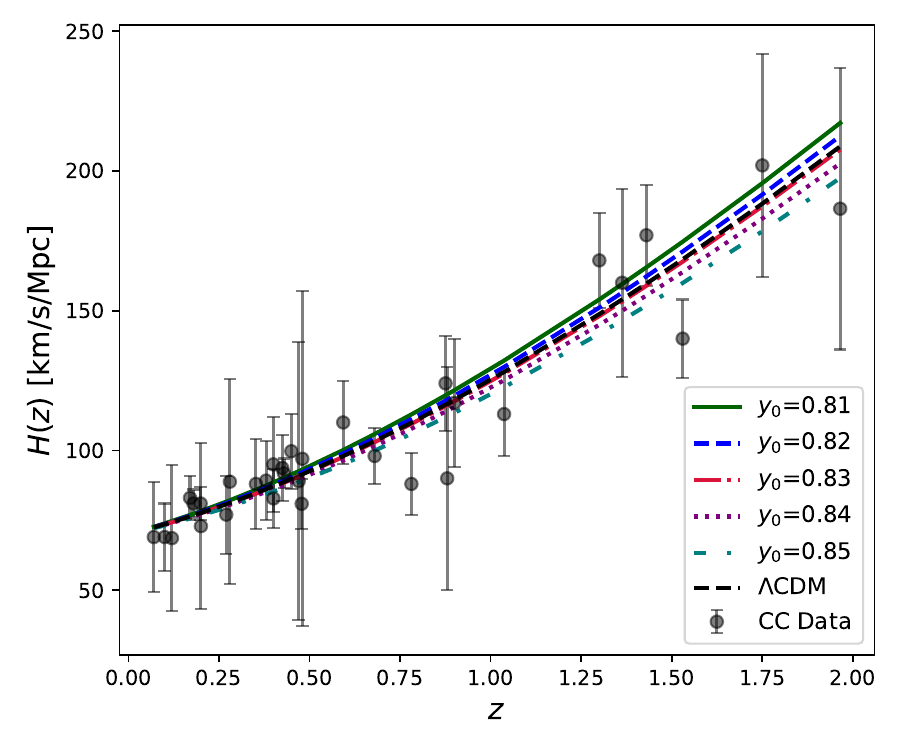} 
    \caption{Evolution of the Hubble parameter \( H(z) \) as a function of redshift \( z \) for different values of the model parameter \( y_0 \) for Model II. The curves correspond to theoretical predictions for \( y_0 \in (0.81,0.85 )\), along with the standard \( \Lambda \)CDM model. The data points with error bars represent cosmic chronometer (CC) measurements of \( H(z) \). The plot illustrates the impact of varying \( y_0 \) on the expansion history and its consistency with observational data. All model curves are generated using the same set of initial conditions and a fixed value of \( H_0 = 70\, \text{km/s/Mpc} \).
}
    \label{ModelHB}
\end{figure}

Model II exhibits a qualitatively similar evolution of the cosmological density parameters $\Omega_{\phi}$, $\Omega_{\psi}$, and $\Omega_{r}$, as well as the total equation-of-state parameter $w_{\text{tot}}$ as those presented in Model I, and shown explicitly in Figs. \ref{fig:ModelB} and \ref{ModelHB}. Initialized deep in the radiation era at $N=-20$, the system reproduces the standard sequence of radiation, matter, and late-time accelerated expansion. At the early time, radiation dominates with $\Omega_r \approx 1$, followed by a prolonged matter-dominated epoch characterized by $w_{\text{tot}} \approx 0$. At late times, the scalar field gradually overtakes the matter sector, driving acceleration with $w_{\text{tot}} \to -1$. The exponential potential ensures a smooth and gradual transition to acceleration while maintaining consistency with Big Bang Nucleosynthesis and present-day observational constraints.

\section{Data Analysis}
\label{sec:data}

In this section, we present the observational data sets to constrain the model parameters through the Markov Chain Monte Carlo (MCMC) method.
	
\begin{itemize}
\item \textbf{CC Data:} This data set comprises 32 model-independent measurements of the Hubble parameter, commonly known as Cosmic Chronometers (CC). It probes the expansion history by using massive, passively evolving galaxies with old stellar populations and minimal star formation, offering reliable estimates of \( H(z) \) across different redshifts~\cite{Vagnozzi:2020dfn,Jimenez:2001gg,Moresco:2015cya}.

\item \textbf{SNeIa Data:} Type Ia supernovae (SNeIa) are widely used as standard candles due to their relatively uniform intrinsic luminosity. This data set provides measurements of the apparent magnitude \( m_{b}(z) \), from which the luminosity distance \( D_{L}(z) \) is inferred via the magnitude–redshift relation

\begin{equation}
			\mu \equiv m-M = 5\log(D_L/\text{Mpc}) + 25 \  ,
		\end{equation}
		where, \(m\) denotes the apparent magnitude of the supernova and \(D_L\) is the luminosity distance: 	
		\begin{equation}
			D_L({z}) = c(1+{z}) \int_0^{{z}} \frac{dz'}{H(z')} \ ,
		\end{equation} 
assuming a flat FLRW metric, and \(c\) is the speed of light in km/s. The model parameters are constrained by minimizing the chi-square ($\chi^2$) likelihood, defined as:
		\begin{equation}
			-2 \ln (\mathcal{L}) = \chi^2 = \rm \Delta D^{T} \mathcal{C}^{-1} \Delta D_j\ ,
		\end{equation}
		where $\rm	\Delta D = \mu_{\rm Obs} - \mu_{\rm Model}$, $C^{-1}$ denotes the inverse combined statistical and systematic covariance matrix of the SNe sample. We use two different SNe datasets, including PantheonPlus \cite{Brout_2022} and DESY5 \cite{DES:2024jxu,DES:2024hip}.
\begin{itemize}         
\item \textbf{PP Data:} This data set refers to the Pantheon+ compilation, which includes 1550 spectroscopically confirmed Type Ia supernovae. The catalog provides 1770 data samples, from which we use the observational column corresponding to the non-SH0ES-calibrated apparent magnitude \( m_{\rm obs} \), i,e. $0.01 \leq z \leq 2.26$. We denote this subset as ``PP'' throughout our analysis.
		
\item \textbf{DESY5 Data:} This data set consists of Type Ia supernovae observations from the Dark Energy Survey five-year sample (DES-SN5YR), comprising 1829 distinct SNe. It includes 194 nearby SNe with redshift \( z < 0.1 \) and 1635 DES SNe. For our analysis, we compute the likelihood using the distance modulus \( \mu \) and the full covariance matrix provided in the data release. We have updated our supernova analysis to incorporate the DES-Dovekie release, which replaces the earlier DESY5 sample. The DES-Dovekie compilation includes 197 low-redshift SNe~Ia and 1623 DES likely SNe~Ia, with improved calibration and systematic control, making it approximately more robust than the previous dataset. The updated sample shows a mild tension with Pantheon+, which we account for in our joint analysis.

% \item \textbf{Roman Data:} This data set consists of 20,824 simulated data points projected from the forthcoming \textit{Roman Space Telescope}, which is expected to significantly enhance cosmological constraints. A central component of the mission is the High Latitude Wide-Area Survey, incorporating both imaging and spectroscopic observations. In particular, the High Latitude Time-Domain Survey will enable the detection and monitoring of thousands of Type Ia supernovae at redshifts up to \( z \approx 3 \).

% Currently, the dataset is generated using the \texttt{SNANA} simulation package~\cite{Kessler:2009yy}, combined with the \texttt{PIPPIN} pipeline manager, under the assumption of a \( \Lambda \)CDM background. The simulation details were obtained through internal discussions within the Roman Supernova Project Infrastructure Team, with further information available in Refs.~\cite{Kessler:2025eib,Hussain:2024yee,Hounsell:2023xds}.

\end{itemize}

\item \textbf{Roman Data:} This data set consists of 20,824 simulated data points projected from the forthcoming Roman Space Telescope, which is expected to significantly enhance cosmological constraints. A central component of the mission is the High Latitude Wide-Area Survey, incorporating both imaging and spectroscopic observations. In particular, the High Latitude Time-Domain Survey will enable the detection and monitoring of thousands of Type Ia supernovae at redshifts up to \( z \approx 3 \) \cite{RomanSupernovaProjectInfrastructureTeam:2025gzk,Kessler:2025eib}.

Currently, the dataset is generated using the \texttt{SNANA} simulation package~\cite{Kessler:2009yy}, combined with the \texttt{PIPPIN} pipeline manager \cite{Hinton:2020yik}, under the assumption of a \( \Lambda \)CDM background. The simulation details were obtained through internal discussions within the Roman Supernova Project Infrastructure Team, with further information available in Refs.~\cite{Hussain:2024yee,Hounsell:2023xds}.

\item \textbf{DESI BAO:} This data set consists of Baryon Acoustic Oscillation (BAO) measurements from the Dark Energy Spectroscopic Instrument (DESI) Data Release II~\cite{DESI:2025zgx}, which extends and improves upon the earlier DR1 results~\cite{DESI:2024mwx,DESI:2019jxc,Moon:2023jgl}. The key observables are the ratios \( \{D_M/r_d, D_H/r_d, D_V/r_d\} \), where \( D_M \) is the comoving angular diameter distance, \( D_H \) the Hubble distance, \( D_V \) the spherically averaged BAO distance, and \( r_d \) the comoving sound horizon at the drag epoch~\cite{eBOSS:2020yzd,DESI:2024mwx}. For this analysis, we treat \(r_d\) as a free parameter.

\item \textbf{CMB Data:} 
We use the compressed Planck 2018 CMB likelihood based on distance priors \cite{Planck:2018vyg,Chen:2018dbv}, which encode the main geometric information of the temperature power spectrum. The observables include the shift parameter $R$, the angular scale $\theta_*$ (replacing $l_A$), and the physical densities $\Omega_b h^2$ and $\Omega_{\rm dm} h^2$, together with their full covariance matrix as reported in \cite{Arendse:2019hev}. 

The shift parameter is defined as
\begin{equation}
R(z_*) = \frac{D_A(z_*) \sqrt{\Omega_{m_0} H_0^2}}{c},
\end{equation}
where $z_*$ is the redshift of photon decoupling and $D_A$ is the angular diameter distance. The angular scale is given by $\theta_* = r_s(z_*)/D_A(z_*)$, with $r_s$ the comoving sound horizon at decoupling. 

In the interacting scenario considered here, the total matter density is $\Omega_{m_0} = \Omega_{\psi_0} + \Omega_{b_0}$, since baryons remain pressureless and separately conserved while the interaction modifies only the dark matter sector. The redshifts $z_d$ and $z_*$ are computed using the Hu–Sugiyama fitting formulas \cite{Hu:1995en}. The likelihood is constructed through
\begin{equation}
\chi^2_{\rm PLA} = (D_i^{\rm obs}-D_i^{\rm th})\, C^{-1}_{ij}\, (D_j^{\rm obs}-D_j^{\rm th}),
\end{equation}
where $C^{-1}_{ij}$ is the inverse covariance matrix of the compressed dataset.

\end{itemize}

In this work, we perform a joint cosmological analysis using CC, DESI BAO measurements, SNeIa, and CMB data. We consider three combinations of observational datasets: \text{CMB + CC + DESI + PP}, \text{CMB + CC + DESI + DESY5}, and \text{CMB + CC + DESI + Roman}. The total likelihood is constructed as $-2\ln \mathcal{L} = \chi^2_{\text{CMB}} + \chi^2_{\text{CC}} + \chi^2_{\text{BAO}} + \chi^2_{\text{SNe}} $ and is used to constrain Models~I and II, characterized by the free parameters \(\{\Omega_{\phi}, z_0, \Omega_b h^2, H_0\}\). Since $\Omega_{\phi0} = x_0^2 + y_0^2$, we treat $\Omega_{\phi0}$ as an independent parameter in the analysis and one can derive $y_0$ respectively. Furthermore, the dark matter density parameter $\Omega_{\psi}$ can be derived from $z_0$. To explore the parameter space and obtain best-fit values through $\chi^2$ minimization, we employ the \texttt{emcee} MCMC sampler~\cite{Foreman-Mackey:2012any}. For comparison, we also analyze the standard $\Lambda$CDM model.  The prior ranges for the free parameters are chosen as uniform distributions: $\Omega_{\phi} \in \mathcal{U}[0.5,1]$, $z_0 \in \mathcal{U}[0,1]$, 
$H_0 \in \mathcal{U}[50,100]\,\text{km\,s}^{-1}\,\text{Mpc}^{-1}$, and $\Omega_b h^2 \in \mathcal{U}[10^{-5},0.1]$. The posterior distributions are analyzed and visualized using the \texttt{GetDist} package~\cite{Lewis:2019xzd}. To evaluate the statistical performance of the proposed models relative to the spatially flat $\Lambda$CDM scenario, we employ two widely used model selection criteria: the Akaike Information Criterion (AIC) and the Bayesian Information Criterion (BIC)~\cite{Akaike:1974vps,Schwarz:1978tpv}, defined as
\begin{align}
    \mathrm{AIC} &= -2 \ln \mathcal{L}_{\text{max}} + 2k, \label{AIC} \\
    \mathrm{BIC} &= -2 \ln \mathcal{L}_{\text{max}} + k \ln N, \label{BIC}
\end{align}
where \( k \) is the number of free parameters, \( N \) is the total number of data points, and \( \mathcal{L}_{\text{max}} \) is the maximum likelihood.

\section{Results and Discussions}
\label{results}

\begin{figure}
    \centering
       \includegraphics[scale = 0.26]{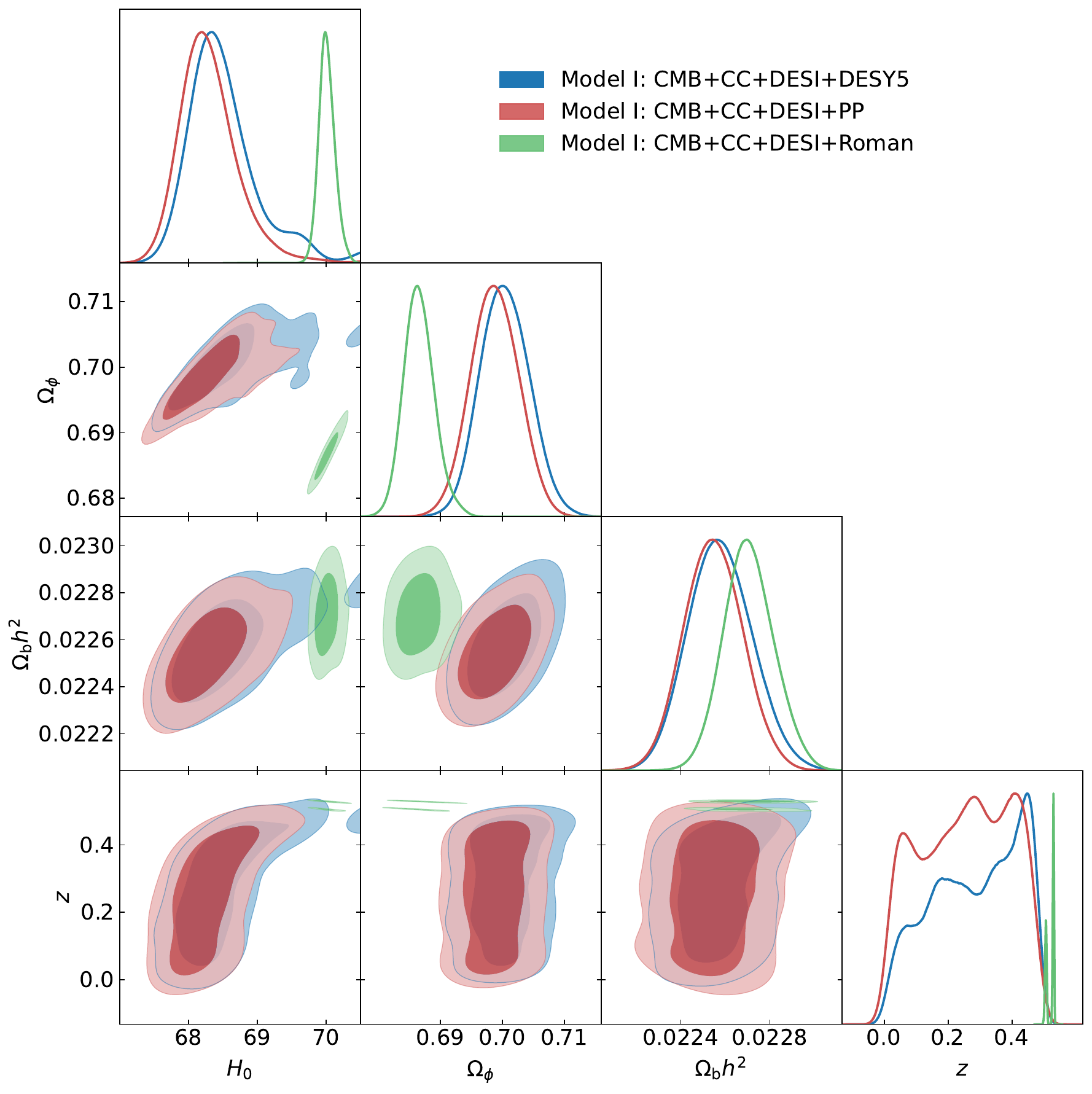} \\ \vspace{0.3cm}  
       \includegraphics[scale = 0.27]{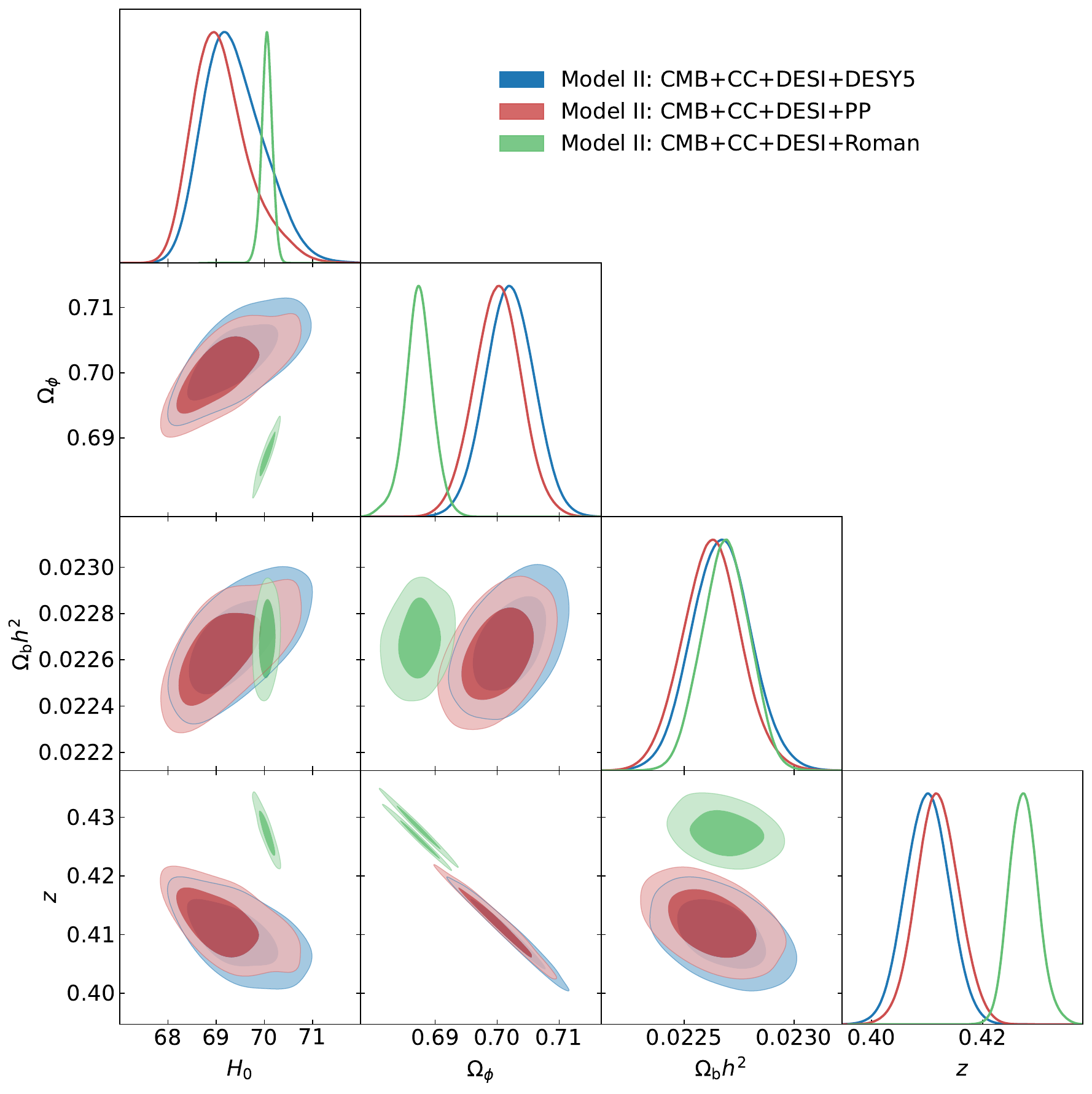}
    \caption{Marginalized one- and two-dimensional posterior distributions (68\% and 95\% CL) for $H_0$, $\Omega_{\phi} = x_0^2 + y_0^2$, $z_0$, and $\Omega_{b0}h^2$ in interacting Models~I and II, derived from the combined CMB+CC+DESI datasets with PP, DESY5 and Roman supernova samples, as shown in the legend.}
    \label{TP-1}
\end{figure}

We present the 68\% confidence level (CL) constraints on the cosmological parameters of Model~I and Model~II, derived from the observational data sets described earlier. The flat \( \Lambda \)CDM model is used as a baseline for comparison. The parameter constraints for both models are summarized in Table~\ref{tab:model_params}, while the one-dimensional marginalized posterior distributions and two-dimensional joint confidence contours are displayed in Figure~\ref{TP-1}.

\begin{table*}[t]
	\begin{tabular} { l  c c c c c}
		\noalign{\vskip 3pt}\hline\noalign{\vskip 1.5pt}\hline\noalign{\vskip 5pt}
        \multicolumn{4}{c}{\textbf{68\% confidence level (CL) constraints for the parameters}} \\ 
\hline
		\multicolumn{1}{c}{\bf Parameters} &  \multicolumn{1}{c}{\bf BASE+PP} &  \multicolumn{1}{c}{\bf BASE+DES} & \multicolumn{1}{c}{\bf BASE+Roman}\\
		
		\noalign{\vskip 3pt}\cline{1-4}\noalign{\vskip 3pt}
		\hline
		\multicolumn{4}{c}{\textbf{Model I}}\\
		\hline
        \hline
		{\boldmath$\Omega_{\phi}  $} & $0.6988 \pm 0.0038$ & $0.70 \pm 0.0039$ & $ 0.6866^{+0.0022}_{-0.0026}$\\
		
		{\boldmath$H_0$} & $   68.33^{+0.28}_{-0.50} $ & $68.56^{+0.26}_{-0.63} $ & $70.01^{+0.10}_{-0.12} $\\
		
		{\boldmath$ z_0   $} & $0.25 \pm 0.14 $ & $0.29^{+0.19}_{-0.13}$ & $0.52^{+0.012}_{-0.016}$ \\
		
		{\boldmath$\Omega_{b}h^2  $} & $0.02255 \pm 0.00013$ & $0.02258 \pm 0.00014 $ & $0.02270 \pm 0.00011$ \\
		\hline
		\hline
		\multicolumn{4}{c}{\textbf{Model II}}\\
		\hline
		\hline
		{\boldmath$\Omega_{\phi}  $} & $0.7001 \pm 0.0038$ & $0.702 \pm 0.0039$ & $0.6873 \pm 0.0023$\\
		
		{\boldmath$H_0$} & $69.11^{+0.43}_{-0.68}$ & $69.37^{+0.51}_{-0.72}$ & $70.05 \pm 0.11$\\

		{\boldmath$z_0 $} & $0.412 \pm 0.0038 $ & $0.410 \pm 0.0039 $ & $0.42 \pm 0.0026 $\\
		
		{\boldmath$\Omega_{b}h^2  $} & $0.02263 \pm 0.00013$ & $0.02267  \pm 0.00013 $ & $0.02269 \pm 0.00011 $ \\
		\hline
		\hline
		\multicolumn{4}{c}{\boldmath \bfseries$\Lambda$CDM}\\
		\hline
		\hline
		{\boldmath$\Omega_{\phi}  $} & $0.6992\pm 0.0038       $ & $0.7009\pm 0.0038 $ & $0.7192\pm 0.0017$\\
		
		{\boldmath$H_0            $} & $68.58\pm 0.30               $ & $68.70\pm 0.30            $ & $71.314\pm 0.099$\\
		
		{\boldmath$\Omega_{b}h^2  $} & $0.02258 \pm 0.00012$ & $0.02260 \pm 0.00013$ &$0.02420^{+0.00011}_{-0.000083}$\\		
	                           
		\hline
		\hline
	\end{tabular}
	\caption{Summary table of cosmological parameter constraints in combination with external datasets and priors, in $\Lambda$CDM and various extended models. Results quoted for all parameters are the marginalized posterior means and 68\% credible intervals in each case. Here, BASE represents the combination CMB+CC+DESI.}
	\label{tab:model_params}
\end{table*}

For consistency, we adopt the combination of CC and DESI DR2 and CMB measurements as our baseline dataset. To evaluate the influence of late-time observational probes, we separately incorporate the Pantheon+ (PP), DESY5, and \textit{Roman} supernova datasets. These combinations enable us to examine how different low-redshift observations affect parameter estimation and the overall behavior of the models.

To prevent numerical instabilities in the background evolution, certain parameters are fixed based on the numerical insights discussed in the previous section. In particular, we fix the initial conditions \( x_0 \) and \( v_0/\Theta_0 \), as well as the model parameters \( \beta \), \( \gamma \), and \( \mu \). Our analysis indicates that these choices have minimal impact on the evolution of the Hubble parameter or energy densities, and the resulting dynamics remain robust across all dataset combinations. We can compute the present-day energy density parameters using the relations $\Omega_{\phi_0} = x_0^2 + y_0^2$ and $\Omega_{\psi_0} = z_0^2 + u_0^2$ evaluated at the best-fit values from the observational constraints. In particular, we analyze the Hubble constant $H_0$ and density parameters as key indicators of the models' viability compared to $\Lambda$CDM. Joint combinations that include PP and DESY yield \( H_0 \approx 68-69~\mathrm{km/s/Mpc} \), while the Roman dataset produces \( H_0 \) around $70~\mathrm{km/s/Mpc}$ . These results indicate a moderate alleviation of the Hubble tension within the interacting dark sector framework, with the extent of the shift depending on the dataset combination. We obtain $\Omega_{\psi} \approx 0.279,\,0.277,$ and $0.29$ for Model~I, and $\Omega_{\psi} \approx 0.276,\,0.275,$ and $0.29$ for Model~II, corresponding to the PP-, DESY5-, and Roman-based combinations, respectively. The Roman dataset favors slightly higher values of $\Omega_{\psi}$, whereas the PP and DESY5 combinations yield comparatively lower estimates. The total matter density parameter can be inferred by combining $\Omega_{\psi}$ with the baryonic contribution $\Omega_b$. Overall, the interacting models remain consistent with the standard cosmological scenario while permitting modest, dataset-dependent variations in the inferred matter content.

\begin{figure}
    \centering
       \includegraphics[scale = 0.42]{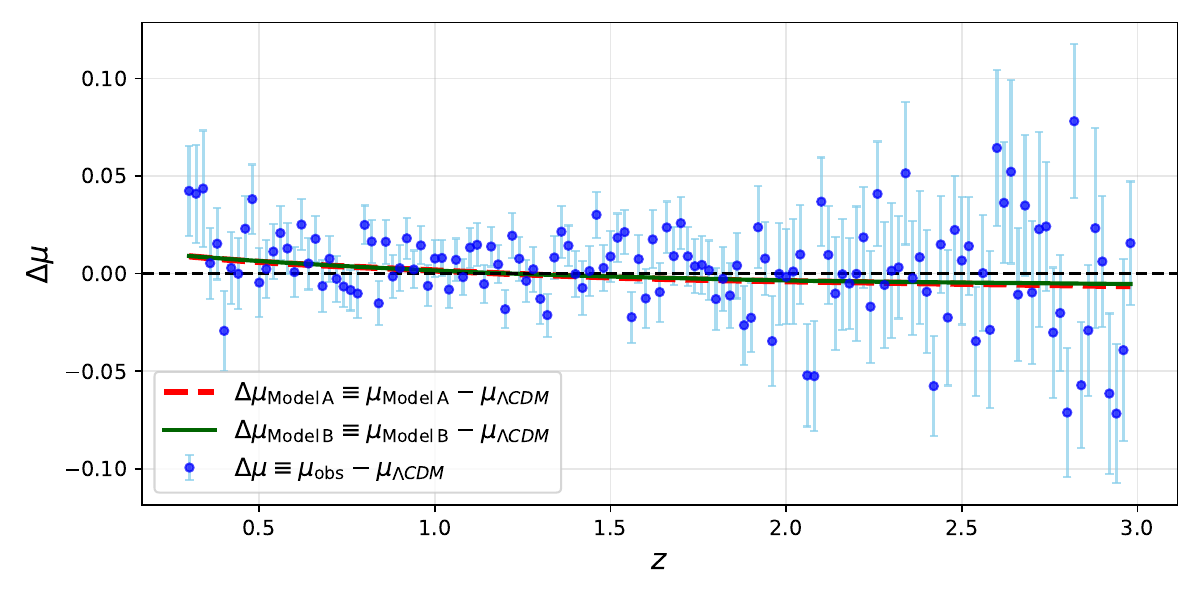} 
     \caption{Residual Hubble diagram showing the difference in distance modulus, \( \Delta\mu = \mu - \mu_{\Lambda \text{CDM}} \), as a function of redshift \( z \) using mock Roman supernova data. The blue points with error bars represent the observed residuals relative to the \(\Lambda\)CDM model.}
    \label{fig:res}
\end{figure}

Figure~\ref{fig:res} shows the residual diagram, where the difference in distance modulus, \( \Delta\mu \equiv \mu - \mu_{\Lambda \text{CDM}} \), is plotted as a function of redshift \( z \). The blue data points with error bars represent the residuals of the mock Roman supernova observations, \( \mu_{\text{obs}} - \mu_{\Lambda \text{CDM}} \), illustrating the spread around the \( \Lambda \)CDM prediction. The red dashed line corresponds to Model~I, and the green solid line to Model~II, with both residuals evaluated relative to \( \Lambda \)CDM. The horizontal black dashed line at \( \Delta\mu = 0 \) indicates perfect agreement with the \( \Lambda \)CDM model.

Both alternative models exhibit small deviations from \( \Lambda \)CDM across the redshift range up to \( z \sim 3 \), which remain well within observational uncertainties. This visual agreement with the mock data suggests that both models are viable alternatives at the background level. To quantitatively assess their relative performance, we employ standard model selection criteria, including \( \Delta\mathrm{AIC} \), \( \Delta\mathrm{BIC} \), and the reduced chi-squared statistic.\footnote{The reduced chi-squared is defined as \( \chi^2_{\mathrm{rd}} \equiv \chi^2_{\mathrm{min}} / \nu \), where \( \nu = N_{\mathrm{obs}} - k \) is the number of degrees of freedom, with \( N_{\mathrm{obs}} \) denoting the total number of observational data points and \( k \) the number of free parameters in the model. A value of \( \chi^2_{\mathrm{rd}} \sim 1 \) typically indicates a good fit, while \( \chi^2_{\mathrm{rd}} \gg 1 \) suggests a poor fit, and \( \chi^2_{\mathrm{rd}} \ll 1 \) may indicate overfitting or excessive model flexibility. For both AIC and BIC, we evaluate the difference relative to the \( \Lambda \)CDM model as \( \Delta IC = IC(\text{model}) - IC(\Lambda \text{CDM}) \).} The results of these statistical comparisons are summarized in Table~\ref{tab:stat}, providing substantial support for the validity of the interacting models.

\begin{table}[]
		\begin{tabular}{|l | c | c | r| r|r|}
			\hline
			Model & AIC & BIC  & $\chi^2_{red} $ & $ \rm \Delta AIC$ & $\rm \Delta BIC$\\
			\hline
			I- BASE+PP & $1483.798$ & $1510.786$ & $0.905$ & $2.848$ & $2.856$ \\
			I - BASE+DESY5 & $1692.08$ & $1719.749$ &  $0.903$ & $1.84$ & $7.37$\\
                I - BASE+Roman & $20984.49$ & $21024.228$ & $1.005$ & $-411.76$ & $-396.852$\\
			\hline
		%	B- DESI+CC & $38.24$  & $46.56$  & $0.830$ &  &  \\
			II- BASE+PP & $1483.797$ & $1510.785$ & $0.905$ & $2.847$ & $2.855$  \\
			II - BASE+DESY5 & $1692.10$ & $1719.756$ &  $0.902$ & $1.86$ & $7.38$ \\
         II - BASE+Roman & $20984.49$ & $21024.229$ & $1.005$ & $-411.76$ & $-396.852$\\
			\hline
			$\Lambda$CDM &  &  & &  &  \\ 
			BASE+PP & $1481.89$ & $1503.49$  & $0.905$ & 0 & 0 \\
			BASE+DESY5 & $1690.24$ & $1712.38$  & $0.904 $& 0 &0 \\
                BASE+ Roman & $21396.25$ & $21420.08$  & $1.025 $& 0 &0 \\
			\hline
			\hline
		\end{tabular}
		\caption{Statistical comparison of the cosmological models based on the reduced chi-squared (\( \chi^2_{\mathrm{rd}} \)), AIC, BIC, and their differences \( \Delta \mathrm{AIC} \), \( \Delta \mathrm{BIC} \). The baseline dataset is the combined CMB+ CC +DESI data.}
		\label{tab:stat}
\end{table}

\section{Conclusions}
\label{conc}

The interaction between dark matter and dark energy has been widely explored and is attracting more and more attention. In this work, we study a non-gravitational interaction within the framework of Einstein–scalar–Gauss–Bonnet (EsGB) gravity, a low-energy limit of the string/M-theory \cite{Gross:1986mw,Bento:1995qc,Ferrara:1996hh,Antoniadis:1997eg}, where dark energy is modeled by  the coupled GB scalar field and dark matter by a fermionic scalar field. The interaction arises naturally from the action, rendering the framework theoretically well-grounded. 

We consider two forms of the GB scalar field potentials: quadratic and exponential, and introduce a coupling between the scalar field and the Gauss–Bonnet term via a general function \( f(\phi) \). The form of this coupling is constrained by recent gravitational wave observations, which place stringent bounds on the propagation speed of tensor modes. This constraint allows the analysis to remain broadly model-independent while remaining consistent with current observational limits.

As a first step, we apply dynamical systems to study the global behavior of the cosmological evolution through numerical techniques. This method offers insights that are largely independent of specific initial conditions. Our results show that both models successfully reproduce the standard sequence of cosmological epochs: radiation-, matter-, and dark energy-dominated epochs, closely tracing the expansion history of the standard  \( \Lambda \)CDM model.

To complement the theoretical analysis, we perform a comprehensive observational study using the most recent data sets, including CMB, CC, SNeIa from Pantheon+, DESY5, and mock Roman survey data, along with BAO measurements from DESI DR2. The results indicate that, for certain combinations of data, the interacting models considered here are mildly favored over the standard \( \Lambda \)CDM model. This finding is particularly interesting in light of the recent DESI DR2 results, which also suggest the possibility of a dynamical dark energy component and a potential departure from the canonical \( \Lambda \)CDM cosmology. 

The statistical analysis indicates that both Model~I and Model~II exhibit a weak preference over $\Lambda$CDM according to the AIC for the PP and DESY5 dataset combinations. This apparent penalization primarily arises from the presence of additional free parameters, which are more strongly disfavored by information criteria such as the BIC due to its explicit dependence on the number of parameters. When these additional parameters are fixed, the corresponding information criteria are substantially reduced and become comparable to those of the standard $\Lambda$CDM model. Hence, although the BIC values are relatively larger, the interacting models cannot be considered significantly disfavored with respect to $\Lambda$CDM for the current datasets. In contrast, for the Roman mock dataset, both interacting models display strong statistical support over $\Lambda$CDM. This suggests that future high-precision surveys may provide meaningful evidence for deviations from the standard cosmological paradigm.

% Additionally, we conduct a comparative analysis together with the standard \( \Lambda \)CDM model using the AIC and BIC as model selection tools. When combined with the Pantheon+ dataset, Models~I and II are nearly indistinguishable from \( \Lambda \)CDM, with both AIC and BIC showing a slight preference from the standard model. However, other dataset combinations reveal a more nuanced picture. In particular, for the \text{DESI+CC+DESY} and \text{DESI+CC+Roman} configurations, the interacting models demonstrate comparable or marginally better performance relative to \( \Lambda \)CDM. These results suggest that the proposed interacting scenarios are competitive with the standard model in accounting for current cosmological observations.

It is noteworthy that the models presented here naturally avoid phantom behavior, owing to the evolution of the Gauss–Bonnet coupling function. When constrained by observational data, this coupling diminishes at low redshifts, effectively suppressing the influence of the GB term in the late Universe. At higher redshifts, although both the GB coupling and interaction terms grow, the parameter space is tightly constrained by stability conditions, ensuring that the equation-of-state remains within the non-phantom regime. This balance is crucial, as previous GB or interacting scalar field models have often faced stability issues, even without crossing into the phantom domain. In this context, the absence of phantom behavior is not a shortcoming, but rather a hallmark of the model’s theoretical robustness and consistency with current observational constraints. Naturally, a complete evaluation of the proposed models requires extending the analysis beyond the background level to include cosmological perturbations. While we anticipate that the inclusion of forthcoming high-precision data will sharpen parameter estimates and further test the robustness of the theory, a detailed perturbative analysis lies is left for future investigations.

\section*{Data Availability}

\noindent Data sharing is not applicable to this article as no datasets were generated during the current study.

\begin{acknowledgments}

\noindent S.A. acknowledges the Japan Society for the Promotion of Science (JSPS) for providing a postdoctoral fellowship during 2024-2026 (JSPS ID No.: P24318). This work of S.A. is also supported by the JSPS KAKENHI grant (Number: 24KF0229). S.H. acknowledges the support of the National Natural Science Foundation of China under Grant No. W2433018 and No. 11675143, and the National Key Research and Development Program of China under Grant No. 2020YFC2201503. B.R. is partially supported as a member of the Roman Supernova Project Infrastructure Team under NASA contract 80NSSC24M0023. This work was completed, in part, with resources provided by the University of Chicago's Research Computing Center. A.W. is partially supported by the US NSF grant, PHY-2308845. We would like to thank the anonymous reviewer for comments and suggestions that helped us to significantly improve our work.
  
\end{acknowledgments}

\appendix

\section{Numerical Analysis of the Autonomous Systems for Models I and II}

In this Appendix, we shall give a detailed analysis of the autonomous systems of Models I $\&$ II considered in our current paper.

In particular, let us first consider Model I, for which the corresponding autonomous system consists of Eqs.(\ref{dyMI1})-(\ref{dyMI8}) and Eq.(\ref{con1}). In the following, rather than determining the critical points analytically, we investigate the stability of the system numerically. Specifically, we evolve the dynamical equations from the deep radiation era to the far future while varying the model parameters within physically motivated ranges. This approach not only enables us to assess the stability properties of the cosmological solutions but also allows us to explore parameter correlations in a systematic manner. To this end, we solve the autonomous system over the interval $N=-20$ to $N=5$. Before performing the numerical evolution, it is essential to impose consistent and physically motivated initial conditions in the early universe.  Here, we therefore describe explicitly the procedure adopted to ensure that the numerical solutions reproduce the standard sequence of cosmological epochs.

\begin{figure*}
    \centering
    \includegraphics[scale = 0.32]{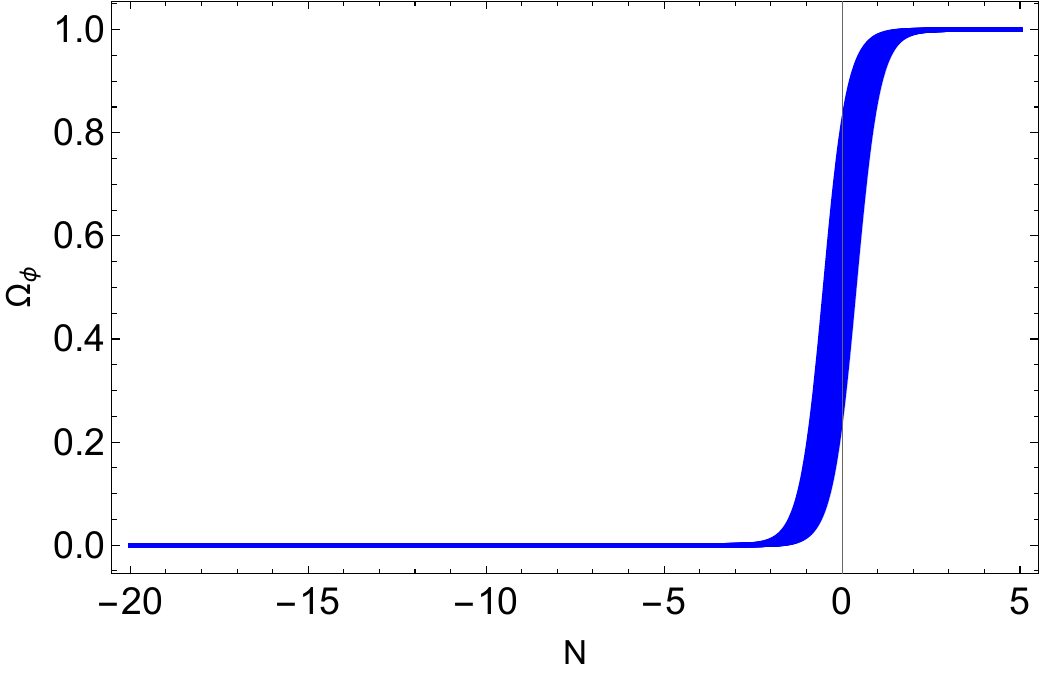} \hspace{0.1cm}    \includegraphics[scale = 0.33]{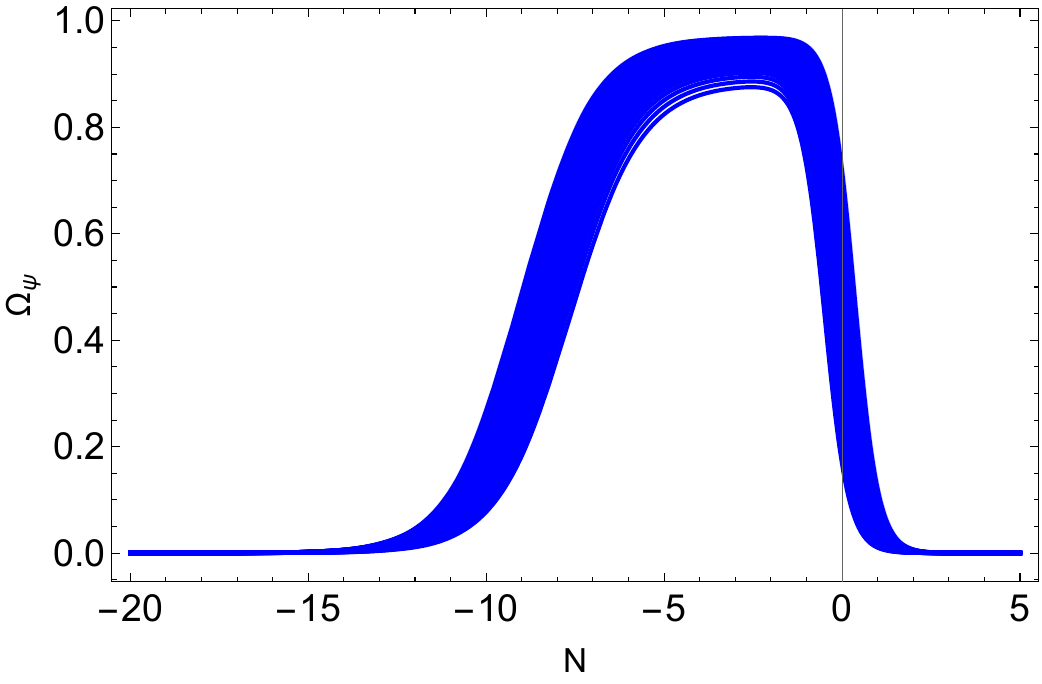}  \hspace{0.1in}
\includegraphics[scale = 0.32]{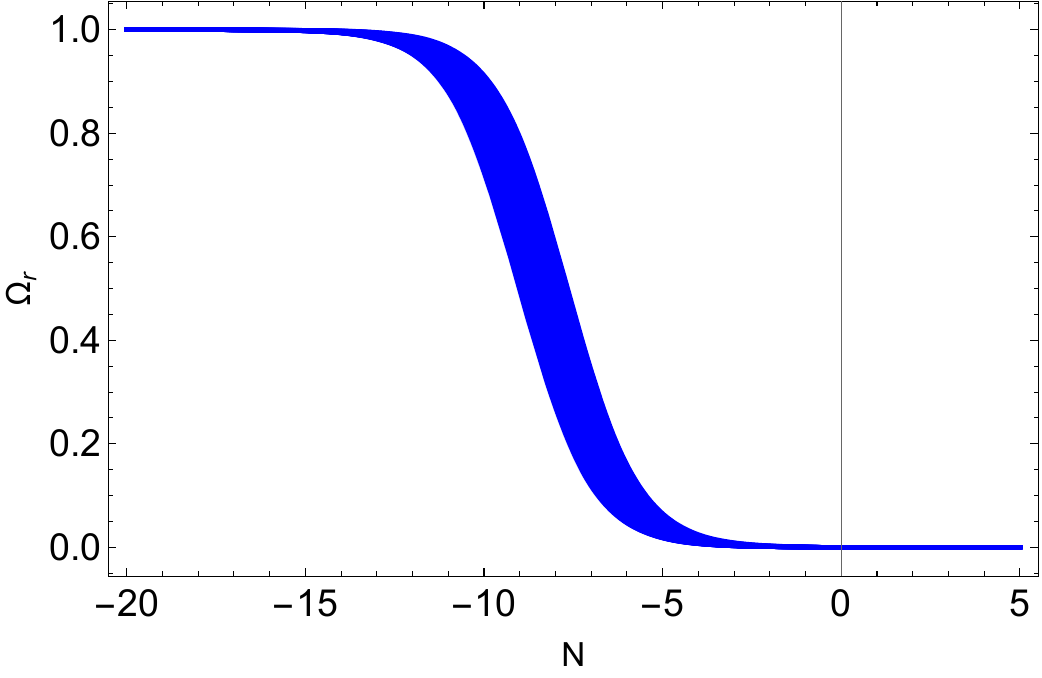} \hspace{0.1in}
\includegraphics[scale = 0.3]{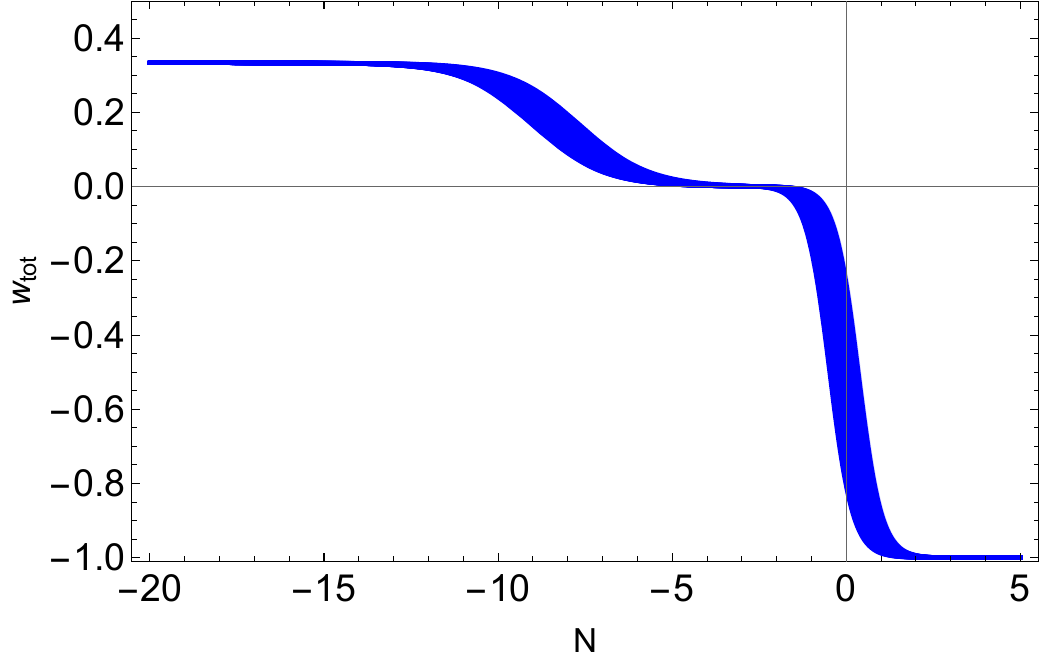} \hspace{0.1in}
\includegraphics[scale = 0.33]{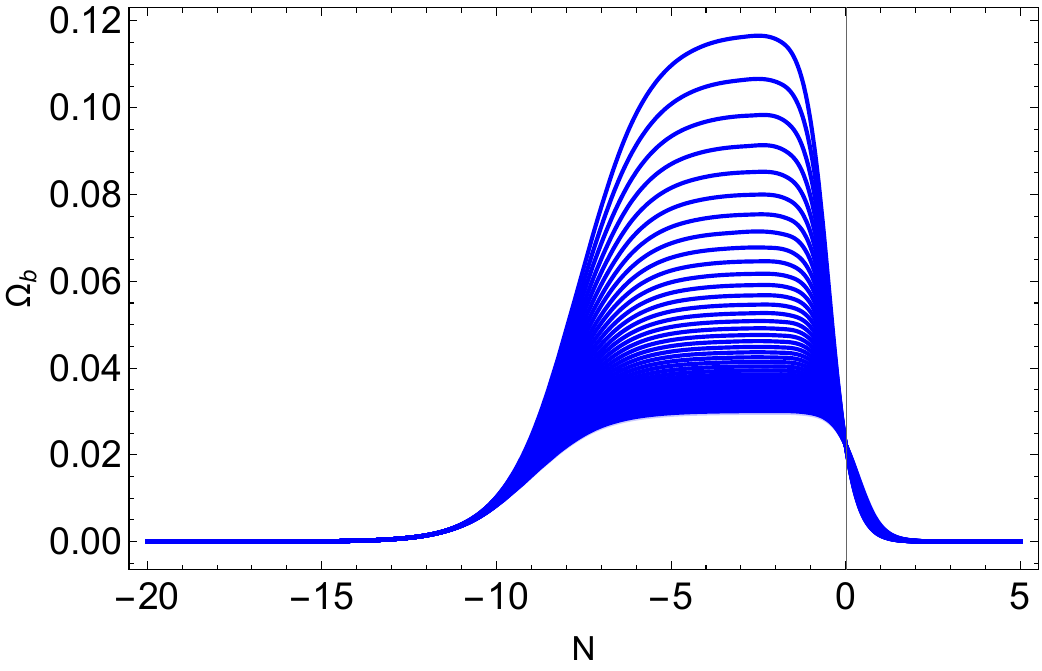} \hspace{0.1in}
    \caption{The evolution of the cosmological parameters 
$\Omega_{\phi}$, $\Omega_{\psi}$, $\Omega_{r}$, 
$\Omega_{b}$ and $w_{\text{tot}}$ for Model~I as functions of $N$. The system is initialized with $\beta = 10^{-8}$, $W_0 = 1$, 
$x_0 = 10^{-13}$, $v_0 = 10^{-35}$, $B_0 = 0.1$, 
$\Omega_{r,0} = 9 \times 10^{-4}$, and $\Omega_{b,0} h^2 = 0.0223$. The initial amplitudes $y_0$ and $z_0$ are randomly generated within the ranges $0.5 \leq y_0 \leq 1.0$ and 
$0.001 \leq z_0 \leq 0.01$, respectively.}
    \label{fig:ModelB2}
\end{figure*}

In this analysis, we first fix the structural parameter $W_0 = 1$ [cf. Eq.(\ref{eq3.15})], since this do not qualitatively alter the dynamical behaviour of the system. The coupling parameter $\beta$ appearing in Eq.(\ref{eq3.14}), which controls the strength of the nonlinear interaction through $\Gamma_2 \propto \beta z^2/u$, is chosen to be $\beta = 10^{-8}$ in order to ensure that the interaction remains perturbative and does not spoil the radiation-dominated era. At the initial time, we assume small values for the scalar kinetic  component and the auxiliary variable, namely $x_0 = 10^{-13}$ and $v_0 = 10^{-35}$, so that the scalar sector is subdominant at early times. The radiation and baryon fractions are fixed to $\Omega_{r,0} = 9 \times 10^{-5}$ and $\Omega_{b,0} = 0.0223$, consistent with observational bounds. We then systematically vary the initial amplitudes $y_0$ and $z_0$, which determine the relative contributions of the scalar potential and effective matter sectors at the present epoch. This procedure allows us to probe the sensitivity of the cosmological evolution to variations in the initial energy partition while keeping the underlying interaction structure fixed. We find that the qualitative sequence of radiation domination, matter domination, and late-time acceleration is preserved across the explored ranges, thereby demonstrating the robustness of the solution. In particular,  in Fig. \ref{fig:ModelB2} we illustrate these robust properties for the representative ranges, $0.5 \leq y_0 \leq 1.0$ and $0.001 \leq z_0 \leq 0.01$.  

For Model II studied in Section III.B, we find qualitatively similar  behaviour under variations of $0.5 \leq y_0 \leq 1.0$ and $0.1 \leq z_0 \leq  0.5$, with the remaining parameters fixed to their corresponding initial values.

 In summary, the standard cosmological sequence of radiation domination, followed by matter domination and late-time acceleration, is preserved  in all cases considered so far. In view of the robustness observed under large variations of the initial amplitudes, when performing the subsequent observational constraints, we can see clearly that  the parameter estimation is not artificially restricted, and that the viable regions of the models are explored comprehensively and physically well accepted.

\bibliography{ref}

\end{document}